\documentclass[aps,amsmath,amssymb,showkeys]{revtex4}
\usepackage[dvips]{graphicx,color}
\usepackage{times}
\usepackage{mathrsfs}
\usepackage{amsmath}
\usepackage{amssymb}
\usepackage{graphicx}
\usepackage{epsfig}
\usepackage{dcolumn}
\usepackage{bm}
\usepackage{xcolor}
\usepackage{hyperref}
\hypersetup{colorlinks=true, citecolor=blue, filecolor=red, urlcolor=cyan}
\newcommand{\bqn}{\begin{eqnarray}}
\newcommand{\eqn}{\end{eqnarray}}
\newcommand{\nb}{\nonumber}
\newcommand{\lb}{\label}

\newcommand{\p}{\partial}
\newcommand{\etal}{{\it et al.\ }}
\newcommand{\be}{\begin{equation}}
\newcommand{\ee}{\end{equation}}

\newcommand{\orcid}[1]{\href{https://orcid.org/#1}{\includegraphics[width=8pt]{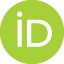}}}

\begin{document}
\title{Weak gravitational lensing and shadow cast by rotating black holes in 
axionic Chern-Simons theory}

\author{Nashiba Parbin \orcid{0000-0002-3570-0117}}
\email[Email: ]{nashibaparbin91@gmail.com}
\affiliation{Department of Physics, Dibrugarh University,
Dibrugarh 786004, Assam, India}

\author{Dhruba Jyoti Gogoi \orcid{0000-0002-4776-8506}}
\email[Email: ]{moloydhruba@yahoo.in}

\affiliation{Department of Physics, Dibrugarh University,
Dibrugarh 786004, Assam, India}

\author{Umananda Dev Goswami  \orcid{0000-0003-0012-7549}}
\email[Email: ]{umananda@dibru.ac.in}

\affiliation{Department of Physics, Dibrugarh University,
Dibrugarh 786004, Assam, India}

\begin{abstract}
We investigate the impact of the axionic coupling parameter on the bending 
angle of light and the shadow cast by slowly rotating black holes in 
Chern-Simons modified gravity. We utilize the Ishihara \etal method to derive 
the deflection angle of light for an observer and source located at finite 
distances from a lens object in an asymptotically flat spacetime, using the 
Gauss-Bonnet theorem. The deflection angle exhibits an increasing trend up to 
a certain point, followed by a decrease as a function of the impact parameter, 
with the presence of the axion matter field causing the observed increase. 
Additionally, we calculate the Einstein ring radius as a direct application 
of the weak deflection angle. We also investigate the effect of the axion 
matter field on the time delay of light and analyze its impact on the shadow 
cast by slowly rotating black holes. Our findings reveal a significant 
effect of the axionic coupling parameter on the black hole's shadow.

\end{abstract}


\maketitle

\section{Introduction}

The bending of light as it passes through the curved spacetime of the 
gravitational field persists as one of the most convenient observational tools 
to understand the spacetime geometry encompassing a strong gravitational source 
\cite{schneider, trimble, renn, valls}. Observed for the first time in $1919$ 
during a solar eclipse \cite{eddington}, the gravitational deflection of light 
led to the first experimental verification of Einstein's theory of General 
Relativity (GR) \cite{einstein}. A noteworthy implementation of gravitational 
bending is the study of weak lensing. The distribution of dark matter in 
galaxies and galaxy clusters, identification of extrasolar planets, etc.\ can 
be revealed by weak gravitational lensing studies. In recent times, weak 
lensing phenomena has become a central topic of research in modern astronomy 
and cosmology.

From the theoretical as well as the observational points of view, the study of 
null geodesics around a black hole plays a crucial role in determining 
gravitational field features of the black hole, such as its gravitational 
lensing and shadow. Gravitational lensing around black holes have been 
investigated in many scenarios such as the Schwarzschild-like black holes 
\cite{virbhadra}, AdS/dS black holes \cite{zhao, rindler}, naked singularity 
and horizonless ultracompact objects \cite{ellis, shaikh}, etc. Gibbons and 
Werner altered the standard perspective by discovering a new geometrical 
method to derive the weak deflection angle using the Gauss-Bonnet theorem 
(GBT) \cite{carmo, bonnet} for the static and asymptotically flat 
spacetimes \cite{gibbons}. In this technique, the integral of the theorem can 
be solved in an infinite region surrounded by the ray of light, and an exact 
form of the deflection angle can be derived. The Gibbons-Werner method was 
then applied in the geometry of a stationary black hole spacetime to obtain 
the deflection angle using a Finsler metric of Randers type 
\cite{werner, saavedra}. In $2016$, Ishihara \etal extended the Gibbons-Werner 
method for the finite-distances \cite{ishihara}. Further extensions were 
carried out by Ono \etal to the axisymmetric spacetimes \cite{ono}. These 
generalizations have been employed by various authors for the stationary black 
holes as well as non-asymptotically flat spacetimes 
\cite{kumar, ghosh, banerjee, jusufi, kimet, sakalli, Kjusufi, izzet, ovgun1, 
ovgun2, ovgun3, takizawa}. Crisnejo and Gallo used the GBT to study 
the deflection angle of massive particles as well as light rays for stationary 
spacetimes in a plasma medium \cite{crisnejo}. Few articles have also reported 
the study of the effect of dark matter on deflection angle using the GBT 
\cite{pantig1, pantig2, javed, ovgun4}.

Despite being the most successful theory, GR appears to be inadequate to 
interpret few observational phenomena. The requirement of missing mass in the 
form of dark matter \cite{bertone, swart, tim, frenk, strigari, will2014} to 
describe the galactic rotation dynamics 
\cite{rubin, young1, harko, gergely, parbin} cannot be explained by GR, 
neither can the accelerated expansion of the Universe 
\cite{reiss, perlmutter}. Hence, many theories have been introduced to modify 
GR \cite{Clifton, wagoner, ronald, capozziello, clifton, oikonomou, odintsov, 
sergei, gogoi1, gogoi_cosmo, nashiba}. One of the interesting modified 
theories of gravity is the Chern-Simons (CS) modified gravity theory 
\cite{dorlis,jackiw}. The CS gravity is a widely familiar four dimensional 
scalar-tensor theory, proposed in Ref.~\cite{jackiw}. This theory owns an 
extra dynamical scalar field nonminimally coupled to the Pontryagin density. 
Several studies have been carried out to explore the physical consequences of 
this theory \cite{konno, shaoqi, tiberiu, popov, cisterna, matsuyama, myung, 
molina, cardoso, corral, doneva}. CS black hole spacetimes are unique 
solutions of this theory, and one of its fundamental characteristics is the 
role played by the scalar field. This dynamical scalar field covers the black 
hole with hair by the process of scalarization. In the black hole solution 
under consideration, the scalar field is the string inspired axion matter 
field \cite{witten}, and thus this black hole is dressed in axionic hair. 

The shadow of black holes is another important optical property to understand
their physical attributes. The studies on shadows of different black holes have
been increased significantly in recent times \cite{wonwoo, gussmann, ovgun5, wei, cunha, wang, 
dastan, jamil, kumar, 1s,shnew02,shnew01,2s,3s,4s,5s,6s,7s,8s,9s,10s,11s,12s,
13s,14s,15s,16s,17s,18s,19s,102-1,102-2,102-3,102-4,102-5,102-6,102-7,102-8,
102-9,102-10,102-11, 102-12,102-13,102-14,102-15}. This is due to the fact of 
the recent release of images of black holes by the Event Horizon Telescope 
(EHT) collaboration \cite{akiyama}. The specific shape of the shadow of a black hole 
depends on the physical properties of the black hole being studied \cite{3s}. 
Therefore, the shadow can be used to extract information about the black 
hole's physical properties. Additionally, shadows can help to differentiate 
between various theories of gravity because they are unique to the physical 
properties of the related black holes \cite{3s, carvalho, ronit23, parbin23}. A significant amount of 
literature has also been devoted to studying the shadow cast by a black hole 
surrounded by dark matter \cite{haroon, salucci, hou, konoplya}.

In this work, we intend to explore 
the impact of CS gravity on gravitational lensing of a slowly rotating black 
hole spacetime. CS gravity has already been investigated in various 
aspects, but for the case of gravitational lensing, we study for the 
first time the effects of this gravity on the deflection angle of light by a 
slowly rotating Kerr-type black hole using the GBT. We shall implement the 
Ishihara \etal method and explore the effect of the axion coupling parameter, 
emerging from the CS gravity, on the deflection angle. It is to be mentioned that
the axion is one of the prime candidates of dark matter (DM) 
\cite{nashiba, barkana, cicoli, witten, carosi, ringwald, kawasaki}, 
and hence, from this study we can understand the effect of DM on the 
deflection angle of black holes. We shall also evaluate the Einstein ring 
radius and study the effect of the coupling parameter on it. In addition to 
this, we shall calculate the time delay of light in the CS gravity theory and 
shall analyze the effect of the axion matter field on the time delay. We shall 
also investigate the impact this coupling parameter leaves on the shadow cast 
by the slowly rotating black hole in CS gravity.

The rest of our paper is organized as follows. In Sec.\ \ref{sec.2}, we briefly
review the field equations related to the axionic CS modified gravity theory
and mention the slowly rotating black hole solution for this theory. 
In Sec.\ \ref{sec.3}, we derive the deflection angle in the spacetime of the 
slowly rotating Kerr-type black hole using the Ishihara \etal method. We 
then analyze the effect of the axion coupling parameter on the deflection 
angle for three different black holes viz.\ SgrA$^*$, M$87^*$ and Cen A. In 
Sec.\ \ref{sec.4}, we derive the Einstein ring radius and depict the effect 
of the coupling parameter. Next, we compute the time delay of light and the 
effect of the axion matter field on the time delay, in Sec.\ref{sec.5}. 
Furthermore, we study the shadow cast by the black hole in Sec.\ \ref{sec.6}. 
Finally, in Sec.\ \ref{sec.7}, we present the summary and conclusions of our 
work. Throughout our work, we use the sign convention ($-, +, +, +$) 
and the unit $G = 1$.

\section{Chern-Simons gravitational theory and slowly rotating black hole 
solution}
\lb{sec.2}
In the CS gravitational theory two independent fields are taken into 
consideration, one is the gravitational field represented by the spacetime
metric $g_{\mu\nu}$ and other is a scalar field of specific nature. So, in 
particular, the action for the axionic CS gravitational theory is given by 
\cite{dorlis}
\be 
\begin{split}
S &= \int d^4 x \sqrt{-g} \left[\frac{R^\mu_\mu}{2\kappa^2} - \frac{1}{2} (\partial_\mu \varphi) (\partial^\mu \varphi) - \eta\, \varphi\, \mathfrak{R}_{CS}\right] \\[5pt] 
 &= \int d^4 x \sqrt{-g} \left[\frac{R^\mu_\mu}{2\kappa^2} - \frac{1}{2} (\partial_\mu \varphi) (\partial^\mu \varphi) \right] 
 - \int d^4 x\, \eta\,\varphi\, \hat{\mathfrak{R}}_{CS},
 \label{eqn.1}
 \end{split}
\ee
where $\kappa$ is the inverse of the reduced Planck mass $M_{Pl}$, 
$g = \det(g_{\mu\nu})$, $R^\mu_\mu = g^{\mu\nu}R_{\mu\nu}$ is the Ricci scalar
corresponding to Ricci tensor $R_{\mu\nu}$ and $\varphi$ is a pseudoscalar 
representing the axion matter field. In this axionic CS gravitational theory 
the axion matter field is considered to be coupled to the Pontryagin density 
term $\mathfrak{R}_{CS}$ \cite{jackiw, yunes}, which in fact is the 
gravitational CS topological term. This term is expressed as
\be
\mathfrak{R}_{CS} = \frac{1}{2} R_{\nu\rho\sigma}^\mu \tilde{R}^{\nu\ \rho\sigma}_{\mu},
\label{eqn.3}
\ee
where $\tilde{R}^{\nu\ \rho\sigma}_{\mu}$ is the dual of the Riemann tensor 
$R_{\nu\rho\sigma}^\mu$ and is defined as $\tilde{R}^{\nu\ \rho\sigma}_{\mu} = \frac{1}{2} \varepsilon^{\rho\sigma\gamma\delta} R^\nu_{\mu\gamma\delta}$
with $\varepsilon^{\rho\sigma\gamma\delta} = 
\hat{\epsilon}^{\rho\sigma\gamma\delta}/ \sqrt{-g}$, the contravariant four 
dimensional Levi-Civita tensor. Thus the CS term is formed by the contraction 
of the Riemann tensor with its dual tensor. $\hat{\mathfrak{R}}_{CS}$ is the 
CS term with the flat Levi-Civita tensor 
$\hat{\epsilon}^{\rho\sigma\gamma\delta}$ and $\eta$ is the axion
coupling parameter to the CS term. This coupling parameter has the dimension
of length and is expressed in terms of the string Regge slope \cite{duncan}
$\alpha^\prime = M_s^{-2}$, where $M_s$ is the string scale, as
$\eta = \sqrt{2/3}\,\alpha^\prime/48\kappa$,
which is of the order $\mathcal{O}(M_{Pl}/M_s^{2})$.

Varying the action \eqref{eqn.1} with respect to the metric $g_{\mu\nu}$ and
also with the axion matter field $\varphi$, the equations of motion can be 
obtained for the theory as \cite{dorlis},
\begin{align}
G_{\mu\nu} &  = \kappa^2 \,T_{\mu\nu}^{\varphi} + 4\,\kappa^2 \eta\, C_{\mu\nu},
\label{eqn.4}\\[5pt]
\square \varphi &  = \eta\, \mathfrak{R}_{CS},
\label{eqn.5}
\end{align}
where $G_{\mu\nu}$ is the usual Einstein tensor, $C_{\mu\nu}$ is the Cotton 
tensor \cite{jackiw}. Here, the energy-momentum tensor $T_{\mu\nu}^\varphi$ 
is expressed as
\be
T_{\mu\nu}^\varphi = \nabla_\mu \varphi \nabla_\nu \varphi - \frac{1}{2} g_{\mu\nu} (\nabla \varphi)^2
\label{eqn.6}
\ee
The Cotton tensor is obtained by varying the term $\varphi\, \mathfrak{R}_{CS}$ 
with respect to the metric $g_{\mu\nu}$ and can be expressed in the following 
form \cite{dorlis, jackiw}:
\begin{equation}
C_{\mu\nu} = - \frac{1}{2} \nabla^\alpha \left[(\nabla^\beta \varphi) \tilde{R}_{\alpha\mu\beta\nu} + (\nabla^\beta \varphi) \tilde{R}_{\alpha\nu\beta\mu}\right].
\label{eqn.7}
\end{equation}

If we take into consideration the static, spherically symmetric metric, then 
the CS gravitational theory is reduced to GR because the Pontryagin density 
term disappears in this case leading to the absence of the axion matter field. 
This axion field is a string-theory inspired theoretical particle field and is 
a perfect candidate for low-mass DM \cite{odintsov, nashiba}. This field being 
a pseudoscalar, it will impose axial symmetry on any kind of spacetime that we 
work on. Hence, this nature of the theory provides convenient means to find
solutions for rotating compact objects and accordingly, here we consider a 
metric ansatz for slowly rotating Kerr-type black holes as given by 
\cite{dorlis}
\be
ds^2 = - A(r)\, dt^2 + B(r) dr^2 + r^2 d\Omega^2 - 2\, r^2 a \sin^2\! \theta\, W(r)\, dt\, d\phi,
\label{eqn.8}
\ee
where $d\Omega^2 = d\theta^2 + \sin^2\!\theta\, d\phi^2$, $a$ is the spin 
parameter of the rotating black hole spacetime and $W(r)$ is the off-diagonal correction 
term representing the possible backreaction in the spacetime. Considering only 
the leading order of the spin parameter $a$, since the black 
hole is assumed to be of slowly rotating, the solutions of Eqs.\ \eqref{eqn.4} 
and \eqref{eqn.5} for the metric ansatz \eqref{eqn.8} provide  
the metric coefficients, similar to vacuum solutions of Einstein's equations 
as $$A(r) = \left(1 - \frac{2M}{r}\right)\;\; \text{and}\;\; 
B(r) = \left(1 - \frac{2M}{r}\right)^{-1}$$ with the off-diagonal correction 
term as given by
\be
W(r) = \frac{2M}{r^3} - \frac{\eta^2 \kappa^2 (189M^2 + 120M r +70r^2)}{14r^8} + \mathcal{O}(\eta^{2n}),
\label{eqn.9}
\ee
where $n$ is a positive integer $\geq 2$. It is to be noted that the slow 
rotation approximation on the black hole solutions implies that black holes 
are of sufficiently large mass $M$, and for such a case the higher order
terms $\mathcal{O}(\eta^{2n})$ contribute as small perturbations to 
Eq.\ \eqref{eqn.9}. However, it needs to be mentioned that although in the 
black hole solutions above the first or lowest order of the spin parameter $a$ 
has been used, in our rest of the work we consider the orders of $a$ more than 
this as per requirement in the analysis of the features of black holes 
involving variable that is the inverse of the radial distance from the center 
of the black holes \cite{dorlis}.     
   
\section{Deflection angle}
\label{sec.3}
\begin{figure}[ht!]
\includegraphics[scale=0.35]{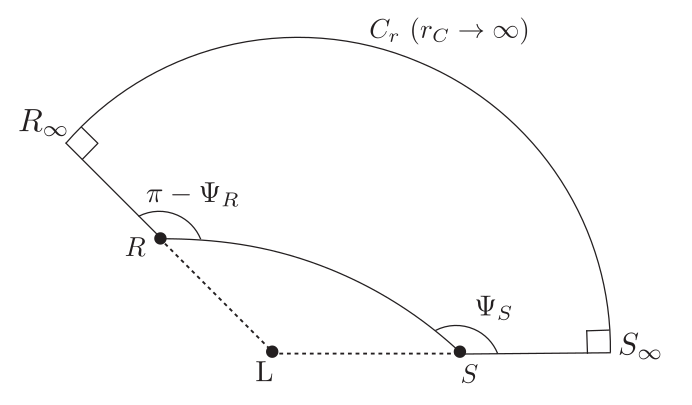}
\caption{Schematic representation for the quadrilateral $\stackrel{\infty}{R}\!\square\!\stackrel{\infty}{S}$ which is embedded in a curved space 
\cite{takizawa}.}
 \label{fig.1}
\end{figure}

In this section, we shall obtain the deflection angle of light in the weak 
field limit of a slowly rotating Kerr-type black hole using GBT approach, 
which was extended by Ishihara \etal\cite{ishihara} in $2016$ as mentioned
earlier. In this approach, the black hole is considered as a lens ($L$), 
which is at a finite distance from the source ($S$) and the receiver ($R$) as 
shown in Fig.\ \ref{fig.1}. For the case of an equatorial plane 
($\theta = \pi/2$), the deflection angle can be expressed as \cite{ishihara, ono}
\be
\hat{\Theta} = \Psi_R - \Psi_S + \Phi_{RS},
\label{eqn.10}
\ee
where $\Psi_R$ and $\Psi_S$ are the angles that are measured at the $R$ and 
the $S$ positions respectively. $\Phi_{RS} = \Phi_R - \Phi_S$ is the 
separation angle between the receiver and the source. Here, $\Phi_R$ and 
$\Phi_S$ are the angular coordinates of the receiver and the source 
respectively. 
The quadrilateral $\stackrel{\infty}{R}\!\square\!\!\stackrel{\infty}{S}$ 
shown in Fig.\ \ref{fig.1} is embedded in a curved space 
$^{(3)}\! \mathcal{M}$, which consists of a spatial curve representing a 
light ray from the $S$ to the $R$, two outgoing radial lines each from the 
$S$ and the $R$, and a circular arc segment $C_r$ having the coordinate 
radius $r_C$ ($r_C \rightarrow \infty$). Using the GBT to this quadrilateral 
$\stackrel{\infty}{R}\!\square\!\!\stackrel{\infty}{S}$, the deflection angle 
\eqref{eqn.10} can also be rewritten as \cite{gibbons}
\be
\hat{\Theta} = - \int\!\! \int_{\stackrel{\infty}{R}\square\stackrel{\infty}{S}} \mathcal{K}\, dS + \int_S^R K_g\, dl,
\label{eqn.11}
\ee
where $\mathcal{K}$ is the Gaussian curvature of the surface of propagation 
of light, $K_g$ is the geodesic curvature of the light curves, $dS$ 
is the infinitesimal area element of the surface and $dl$ is the infinitesimal 
line element of the arc. It is to be mentioned that for the prograde motion of 
photons $dl>0$ and for the retrograde motion $dl<0$.

Thus to obtain the deflection angle for our considered black hole metric 
\eqref{eqn.8}, first we have to study the Gaussian curvature $\mathcal{K}$ of
the propagating light and then calculate its quadrilateral surface integral.  
For this we rewrite the metric \eqref{eqn.8} for the null geodesics 
($ds^2 = 0$) (see Sec.\ \ref{sec.6}) to obtain in the 
form \cite{ishihara, ono}:
\be
dt = \pm\, \sqrt{\zeta_{ij} dx^i dx^j} + \beta_i dx^i,
\label{eqn.12}
\ee
where $\zeta_{ij}$ defines the optical metric and $\beta_i$ represents the 
one-form, which are given by 
\begin{align}
\zeta_{ij} dx^i dx^j & = \frac{dr^2}{\left(1 - \frac{2M}{r}\right)^2} + 
\frac{r^2 d\theta^2}{\left(1 - \frac{2M}{r}\right)} 
+ \frac{r^2 \sin^2 \theta}{\left(1 - \frac{2M}{r}\right)} \nb \\[5pt]
& \times \Biggl\{1 + \frac{r^2 a^2 \sin^2\theta}{\left(1 - \frac{2M}{r}\right)} \left[\frac{2M}{r^3} - 
\frac{\eta^2 \kappa^2 (189 M^2 + 120 M r + 70 r^2)}{14 r^8}\right]^2 \Biggr\}\,d\phi^2
\label{eqn.13}
\end{align}
and
\be
\beta_i dx^i = - \frac{r^2 a \sin^2 \theta}{\left(1 - \frac{2M}{r}\right)} \Biggl[\frac{2M}{r^3} - 
\frac{\eta^2 \kappa^2 (189 M^2 + 120 M r + 70 r^2)}{14 r^8} \Biggr] d\phi.
\label{eqn.14}
\ee
With this optical metric the Gaussian curvature of propagating light is 
defined as \cite{gibbons}
\be
\mathcal{K} = \frac{^{(3)\!} R_{r\phi r\phi}}{\zeta} = \frac{1}{\sqrt{\zeta}} \left[\frac{\p}{\p \phi} \left(\frac{\sqrt{\zeta}}{\zeta_{rr}}\ ^{(3)\!}\zeta_{rr}^\phi\right) - \frac{\p}{\p \phi} \left(\frac{\sqrt{\zeta}}{\zeta_{rr}} \ ^{(3)\!}\zeta_{r\phi}^\phi\right) \right],
\label{eqn.15}
\ee
where $\zeta \equiv \det(\zeta_{ij})$. For the slowly rotating black hole 
in CS gravitational theory, $\mathcal{K}$ is computed for propagation of light 
in the equatorial plane as
\be
\mathcal{K} = - \frac{2 M}{r^3} + \frac{3 M^2}{r^4} - \frac{24 a^2 M^2}{r^6} + \frac{16 a^2 M^3}{r^7} + \mathcal{O}\Big(\frac{1}{r^{8}}, \frac{a^2}{r^{8}}, a^3\Big).
\label{eqn.16}
\ee
Here terms $\mathcal{O}(1/{r^{8}}, a^2/r^{8}, a^3)$ are 
found to be very small in comparison to the retaining terms. Using this 
expression \eqref{eqn.16} the surface integral of the Gaussian curvature
$\mathcal{K}$ over the closed quadrilateral can be computed from the equation 
\cite{ono},
\be
- \int \!\! \int_{\stackrel{\infty}{R}\square\stackrel{\infty}{S}} \mathcal{K}\, dS = \int_{\phi_S}^{\phi_R}\!\! \int_{\infty}^{r_{ps}} \mathcal{K} \sqrt{\zeta}\,dr\, d\phi,
\label{eqn.17}
\ee
where $r_{ps}$ is the radius of the photon sphere (see Sec.\ \ref{sec.6}). 
Further, we need to study the photon orbit equation in the equatorial plane 
for the metric \eqref{eqn.8} to derive the integrals in Eq.\ \eqref{eqn.17}.
For this one should note that there are two constants of motion along the 
orbit in the equatorial plane associated with two Killing vectors 
$\p/\p \tau$ and $\p/\p \phi$ as given by 
\cite{ishihara, ono}
\begin{align}
\mathcal{E} & = A(r) \dot{t} + r^2 a W(r) \dot{\phi},
\label{eqn.18}\\[5pt]
L_z & = r^2 \dot{\phi} - r^2 a W(r)\, \dot{t},
\label{eqn.19}
\end{align}
where the dot over the variables denotes the derivative with respect to the 
affine parameter $\tau$. From these two constants of motion, we obtain the 
impact parameter from the usual definition as
\be
\Upsilon \equiv \frac{L_z}{\mathcal{E}} 
 = \frac{r^2 \dot{\phi} - r^2 a W(r)\, \dot{t}}{A(r) \dot{t} + r^2 a W(r) \dot{\phi}}.
\label{eqn.20}
\ee
The null geodesic condition, $ds^2 = 0$ in Eq.\ \eqref{eqn.8} leads to the 
photon or light orbit equation in the equatorial plane, which is given by
\be
\left(\frac{dr}{d\phi}\right)^{\!2} = \frac{A(r) r^2 + \big(r^2 a W(r)\big)^2}{B(r)}\left[\frac{r^2 - 2r^2 a W(r) \Upsilon - A(r) \Upsilon^2}{\big(r^2 a W(r) + A(r)\Upsilon\big)^2}\right].
\label{eqn.21}
\ee
To make the line integration limit of Eq.\ \eqref{eqn.17} finite, we change the 
variable $r$ to $u$ by $r \equiv \frac{1}{u}$. Accordingly the above equation 
can be rewritten as
\be
\left(\frac{du}{d\phi}\right)^2 = F(u),
\label{eqn.22}
\ee
where
\be
F(u) = \frac{u^2 (1 - 2M u) \left[(1 - 2M u)u^2 + a^2 W^2\right] 
\left[1 - (1 - 2M u)\Upsilon^2 u^2 - 2a W \Upsilon\right]}{\big[a W + (1 - 2M u) \Upsilon u^2\big]^2}.
\label{eqn.23}
\ee
In the weak field limit as well as in the slow rotation approximation, the 
iterative solution of Eq.\ (\ref{eqn.22}) is obtained as \cite{ono}
\be
u = \frac{\sin\phi}{\Upsilon} + \frac{M(1 + \cos^2\phi)}{\Upsilon^2} - \frac{2aM}{\Upsilon^3}
\label{eqn.24}
\ee
and hence, we can rewrite Eq.\ \eqref{eqn.17} as
\be
- \int\!\! \int_{\stackrel{\infty}{R}\square\stackrel{\infty}{S}} \mathcal{K} dS = \int_{\phi_s}^{\phi_R} \!\!
\int_0^u - \,\frac{\mathcal{K}\sqrt{\zeta}}{u^2}\, du\, d\phi.
\label{eqn.25}
\ee
Thus for the slowly rotating black hole metric \eqref{eqn.8} in axionic CS 
gravity, the above equation can be integrated out using Eqs.\ \eqref{eqn.13},
\eqref{eqn.16} and \eqref{eqn.24} as
\begin{align}
- \int\!\! \int_{\stackrel{\infty}{R}\square\stackrel{\infty}{S}} &
\mathcal{K}\, dS = \bigg(\frac{2M}{\Upsilon} + \frac{21 M^3}{4\Upsilon^3} 
- \frac{6a M^3}{\Upsilon^4} + \frac{57a^2 M^3}{2\Upsilon^5} + 
\frac{767M^5}{32\Upsilon^5} - \frac{195a M^5}{4\Upsilon^6} + 
\frac{10615 a^2 M^5}{32\Upsilon^7} \nb \\[5pt] 
& + \frac{66705M^7}{512\Upsilon^7} - \frac{11985a M^7}{32\Upsilon^8}\bigg) \bigg(\sqrt{1 - \Upsilon^2 u_R^2} - 
\sqrt{1 - \Upsilon^2 u_S^2}\bigg) + \bigg(\frac{M^3}{12\Upsilon^3} - 
\frac{11a^2 M^3}{4\Upsilon^5}\nb \\[5pt] 
& + \frac{3M^5}{64\Upsilon^5} + \frac{aM^5}{12\Upsilon^6} -  
\frac{1081 a^2 M^5}{96\Upsilon^7} - \frac{1317M^7}{512\Upsilon^7} + 
\frac{475a M^7}{64\Upsilon^8}\bigg) \Big[\big(1 - \Upsilon^2 u_R^2\big)^{3/2} - 
\big(1 - \Upsilon^2 u_S^2\big)^{3/2}\Big]\nb \\[5pt] 
& - \bigg(\frac{3a^2 M^3}{20\Upsilon^5} + \frac{M^5}{64\Upsilon^5} 
+ \frac{93a^2 M^5}{160\Upsilon^7} + \frac{143M^7}{512\Upsilon^7} - 
\frac{99a M^7}{320\Upsilon^8}\bigg) \Big[\big(1 - \Upsilon^2 u_R^2\big)^{5/2} 
- \big(1 - \Upsilon^2 u_S^2\big)^{5/2}\Big]\nb \\[5pt] 
& + \bigg(\frac{5a^2 M^5}{1568\Upsilon^7} - \frac{11M^7}{3584\Upsilon^7}\bigg) 
\Big[\big(1 - \Upsilon^2 u_R^2\big)^{7/2} - 
\big(1 - \Upsilon^2 u_S^2\big)^{7/2}\Big] + \bigg(- \frac{M^2}{4\Upsilon^2} + 
\frac{3a^2 M^2}{\Upsilon^4} \nb \\[5pt] 
& + \frac{37M^4}{16\Upsilon^4} - \frac{3a M^4}{\Upsilon^5} + 
\frac{1925a^2 M^4}{32\Upsilon^6} + \frac{10293M^6}{512\Upsilon^6} - 
\frac{183aM^6}{4\Upsilon^7}\bigg) \bigg(\Upsilon u_R \sqrt{1 - 
\Upsilon^2 u_R^2}\nb \\[5pt] 
& + \Upsilon u_S \sqrt{1 - \Upsilon^2 u_S^2}\bigg) + 
\bigg(\frac{15M^2}{4\Upsilon^2} - \frac{4aM^2}{\Upsilon^3} + 
\frac{9a^2 M^2}{4\Upsilon^4} + \frac{543M^4}{64\Upsilon^4} - 
\frac{15aM^4}{\Upsilon^5} + \frac{3317a^2 M^4}{48\Upsilon^6}\nb \\[5pt]
& \frac{9591M^6}{256\Upsilon^6} - \frac{1521aM^6}{16\Upsilon^7}\bigg) 
\big[\pi - \arcsin(\Upsilon u_R) - \arcsin(\Upsilon u_S)\big] + 
\mathcal{O}\Big(\frac{1}{\Upsilon^8}, \frac{a^2}{\Upsilon^8}, a^3\Big).
\label{eqn.26}
\end{align}
In the above expression, we use $u_R$ as the reciprocal of the distance of the 
receiver from the black hole and $u_S$ as that of the source from the black 
hole. We also use $\cos \phi_R = - \sqrt{1 - \Upsilon^2 u_R^2}$ and 
$\cos \phi_S = \sqrt{1 - \Upsilon^2 u_S^2}$ \cite{ono}. In the far distance 
limit, $u_R \rightarrow 0$ and $u_S \rightarrow 0$, this Eq.\ \eqref{eqn.26} 
takes the form:
\begin{align}
- \int \!\!\int_{\stackrel{\infty}{R}\square\stackrel{\infty}{S}} 
\mathcal{K}\, dS \approx &\, \frac{4M}{\Upsilon} + 
\frac{15\pi M^2}{4\Upsilon^2} - \frac{4\pi a M^2}{\Upsilon^3} 
+ \frac{32M^3}{3\Upsilon^3} + \frac{9\pi a^2 M^2}{4\Upsilon^4}\nb \\[5pt] 
& - \frac{12a M^3}{\Upsilon^4} + \frac{256a^2 M^3}{\Upsilon^5} + 
\mathcal{O}\Big(\frac{1}{\Upsilon^6}, \frac{a^2}{\Upsilon^6}, a^3\Big).
\label{eqn.26a}
\end{align}

Next, we have to study the geodesic curvature of light and then calculate
its path integral. In the equatorial plane ($\theta = \pi/2$), the geodesic 
curvature in the manifold $^{(3)\!}\mathcal{M}$ can be expressed as
\be
K_g = - \frac{1}{\sqrt{\zeta \zeta^{\theta\theta}}}\, \beta_{\phi,r},
\label{eqn.27}
\ee
which for the slowly rotating black hole metric (\ref{eqn.8}) yields,
\be
K_g = -\frac{2 a M}{r^3}-\frac{2 a M^2}{r^4}-\frac{3 a M^3}{r^5} + 
\mathcal{O}\Big(\frac{a}{r^{6}}, a^3\Big).
\label{eqn.28}
\ee
To derive the path integral of the geodesic curvature, let us consider 
that the coordinate system is centered at the lens position. For such a case, 
the light curve can be approximated with, $r = \Upsilon/\sin\!\vartheta$ and 
$l = \Upsilon \tan\vartheta$ \cite{ono}. Using these two relations in 
Eq.\ \eqref{eqn.28}, the path integral of the geodesic curvature can be
written as
\be
\int_S^R K_g dl = \int_{\phi_S}^{\phi_R} \bigg[-\, \frac{2aM}{\Upsilon^2} \sin\!\vartheta - \frac{2aM^2}{\Upsilon^3} \sin^2\!\vartheta - \frac{3aM^3}{\Upsilon^4} \sin^3\!\vartheta + \mathcal{O}\Big(\frac{a}{\Upsilon^5}, a^3\Big)\bigg] d\vartheta.
\label{eqn.29}
\ee
The evaluation of integration of this equation leads to its explicit form as
\begin{align}
\int_S^R K_g dl = &\, - \bigg(\frac{2aM}{\Upsilon^2} + 
\frac{9aM^3}{4\Upsilon^4}\bigg) \bigg(\sqrt{1 - \Upsilon^2 u_R^2} + 
\sqrt{1 - \Upsilon^2 u_S^2}\bigg)\nb \\[5pt] 
& - \frac{aM^2}{\Upsilon^3}\bigg(\Upsilon u_R \sqrt{1 - \Upsilon^2 u_R^2} + 
\Upsilon u_S \sqrt{1 - \Upsilon^2 u_S^2}\bigg)\nb \\[5pt] 
& - \frac{3aM^3}{4\Upsilon^4}\left(\Upsilon^2 u_R^2 \sqrt{1 - \Upsilon^2 u_R^2} + \Upsilon^2 u_S^2 \sqrt{1 - \Upsilon^2 u_S^2}\right)\nb\\[5pt] 
& + \frac{aM^3}{4\Upsilon^4}\Big[\big(1 - \Upsilon^2 u_R^2\big)^{3/2} - 
\big(1 - \Upsilon^2 u_S^2\big)^{3/2}\Big]\nb \\[5pt]
& - \frac{aM^2}{\Upsilon^3}\big[\pi - \arcsin(\Upsilon u_R) - 
\arcsin(\Upsilon u_S)\big] + \mathcal{O}\Big(\frac{a}{\Upsilon^5}, a^3\Big),
\end{align}
where we use $\cos\phi_R = - \sqrt{1 - \Upsilon^2 u_R^2}$ and 
$\cos\phi_S = \sqrt{1 - \Upsilon^2 u_S^2}$ \cite{ono}. Moreover, in this 
derivation we consider the prograte motion ($dl>0$) wherein the orbital motion 
of photons is in the same direction with that of spin of the black hole.
In far distance limit, $u_R \rightarrow 0$ and $u_S \rightarrow 0$, this 
equation becomes,
\be
\int_S^R K_g dl \approx - \frac{4aM}{\Upsilon^2} - \frac{\pi aM^2}{\Upsilon^3} - \frac{9aM^3}{2\Upsilon^4} + \frac{aM^3}{2\Upsilon^4} + \mathcal{O}\Big(\frac{a}{\Upsilon^5}, a^3\Big).
\label{eqn.29a}
\ee
 
Hence, we arrive at the expression for the deflection angle of light by a 
slowly rotating black hole in the axionic CS gravity theory by combining 
Eqs.\ \eqref{eqn.26a} and \eqref{eqn.29a} in the asymptotically far distance 
limit, $u_R \rightarrow 0$ and $u_S \rightarrow 0$, which is obtained as
\begin{align}
\hat{\Theta} \approx &\, \frac{4M}{\Upsilon} - \frac{4aM}{\Upsilon^2} + \frac{15\pi M^2}{4\Upsilon^2} - \frac{5\pi aM^2}{\Upsilon^3} + \frac{32M^3}{3\Upsilon^3}
\nb\\[5pt]
& + \frac{9a^2 M^2 \pi}{4\Upsilon^4} - \frac{41aM^3}{2\Upsilon^4} + \frac{256a^2 M^3}{5\Upsilon^5}.
\label{eqn.30}
\end{align}
It is to be noted that in the limiting case of $a = 0$ and $\eta =0$, we 
arrive at the deflection angle for a Schwarzschild black hole as given by 
\cite{virbhadra}
\be
\hat{\Theta} \approx \frac{4M}{\Upsilon_s} + \frac{15\pi M^2}{4\Upsilon_s^2} + \frac{32M^3}{3\Upsilon_s^3},
\label{eqn.31}
\ee
where $\Upsilon_s = r^2 \dot{\phi}/A(r) \dot{t}$ is the impact parameter for
a Schwarzschild black hole. 

\begin{figure}[ht!]
\includegraphics[scale=0.29]{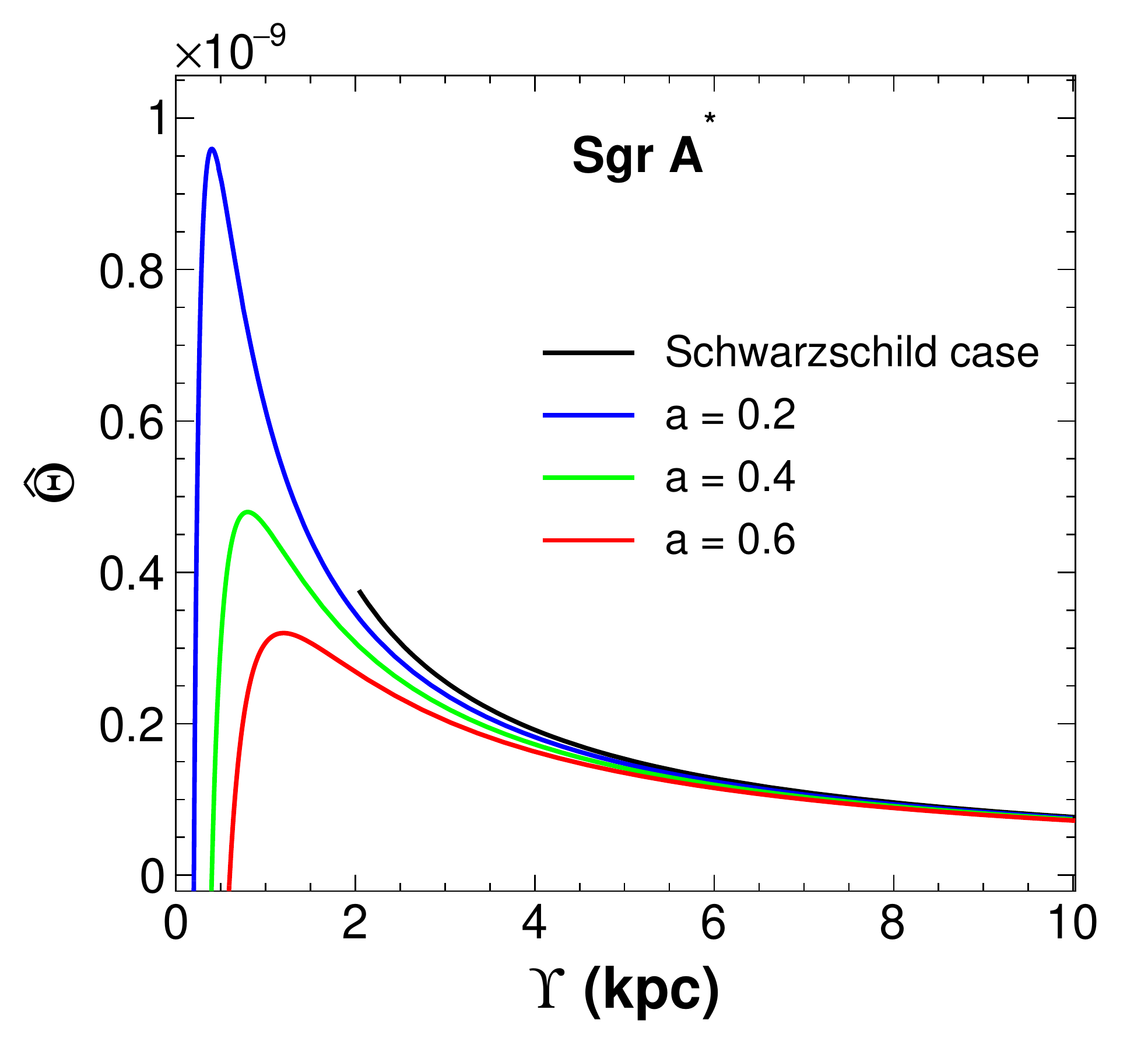}\hspace{1mm}
\includegraphics[scale=0.29]{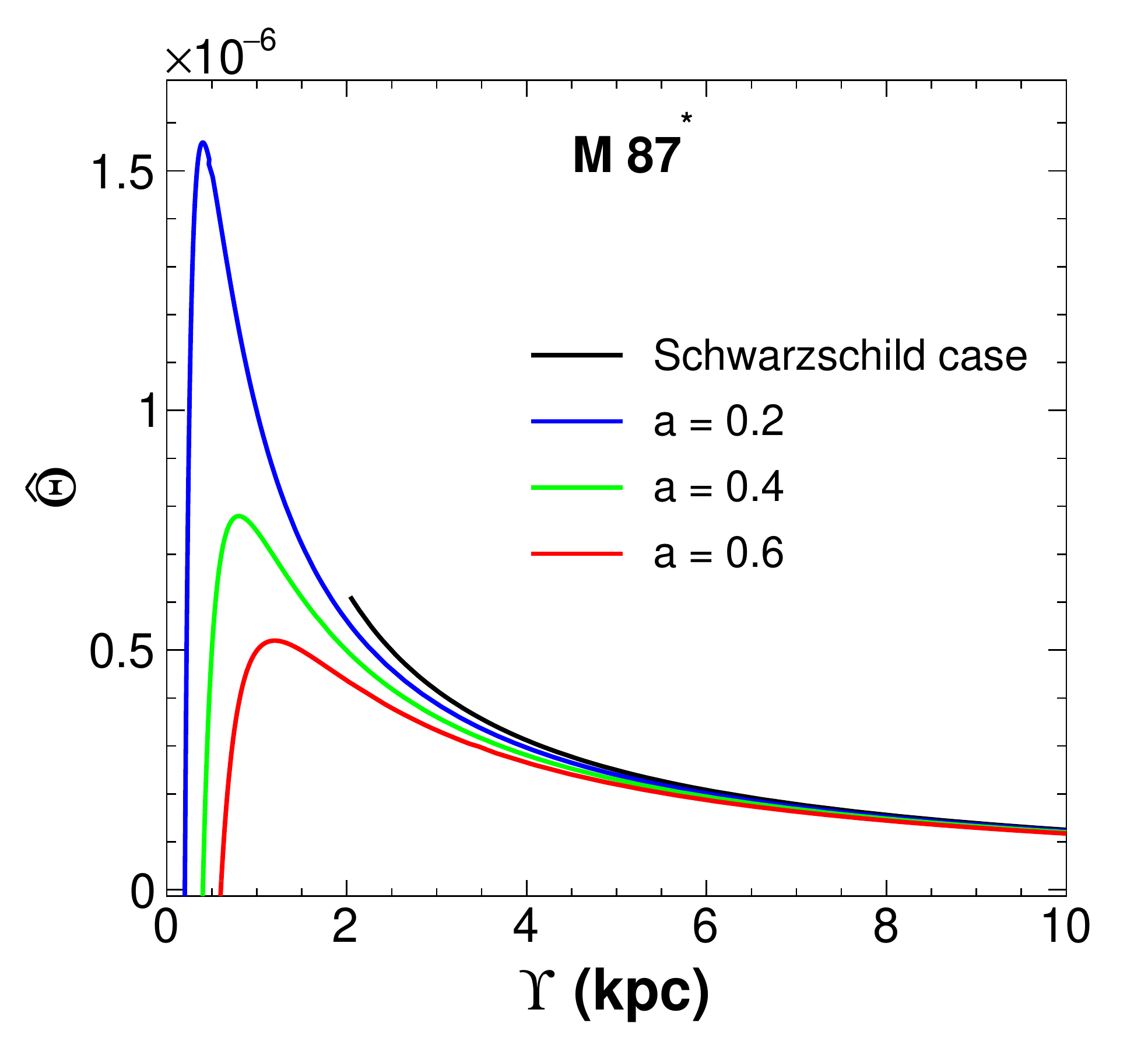}\hspace{1mm}
\includegraphics[scale=0.29]{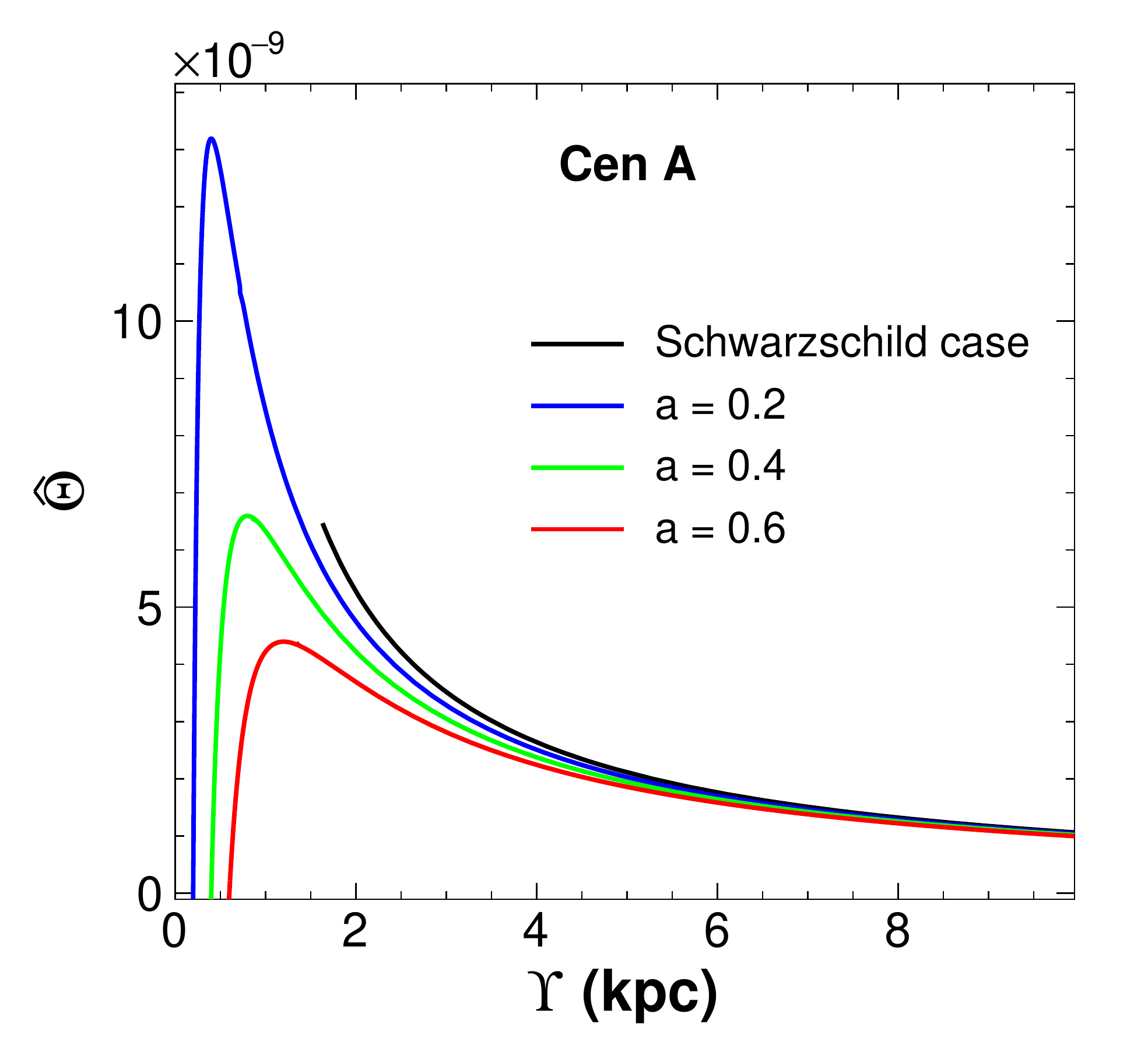}\hspace{1mm}
\caption{Deflection angle as a function of the impact parameter for the three 
black holes, SgrA$^*$, M$87^*$ and CenA with $a = 0.2, 0.4, 0.6$.}
 \label{fig.2}
\end{figure}

Here, we explore the deflection angle for three different black holes, viz.\ 
SgrA$^*$, M$87^*$ and Centaurus A (Cen A) with the respective masses of 
$4 \times 10^6 M_\odot$, $6.4 \times 10^9 M_\odot$ and 
$5.5 \times 10^7 M_\odot$, and study the effect of the axion coupling 
parameter on the deflection angle. The axion coupling parameter depends on 
the string scale $M_s$, which depends on the Planck mass $M_{Pl}$ as 
$M_{Pl} \gtrsim M_s \gtrsim 10^{-3} M_{Pl}$ \cite{basilakos}. In 
Fig.~\ref{fig.2}, we depict the variation of deflection angle $\hat{\Theta}$ 
given by Eq.\ \eqref{eqn.30} as a function of the impact parameter for the 
three black holes and compare our results with the deflection angle 
\eqref{eqn.31} for the corresponding Schwarzschild black holes. For each black 
hole, we see a similar pattern in the behaviour of the deflection angle. It 
can be seen that for different values of the spin parameter $a$, the 
deflection angle increases abruptly upto a certain point for small impact 
parameter values and then decreases in a similar manner as the deflection 
angle for the Schwarzschild case with the increasing values of the impact
parameter. For all the three black holes, it is seen that as the value of the 
spin parameter $a$ grows, the peak of the deflection angle decreases. However, 
for all spin parameter values, the deflection angle finally overlaps with that 
for the Schwarzschild case. It can be said from our results that the increase 
in the deflection angle for small impact parameter value can be due to the 
presence of axion hair of the black hole in the CS modified gravity. 

\section{Einstein ring}
\label{sec.4}
The Einstein ring is an interesting manifestation of the phenomenon of 
gravitational lensing (GL). The shape and behaviour of an Einstein ring 
depends on the position of the lens with respect to the source as well as the 
DM distribution around the lens. A complete Einstein ring is formed when the 
source is perfectly positioned behind the lens. Here, we intend to study the 
effect of the axion coupling parameter $\eta$ on the Einstein ring. To begin 
with, let us define 
$D_{RS}$ as the distance between the source and the receiver, $D_{LS}$ as the 
distance between the lensing object and the source, and $D_{RL}$ as the 
distance between the receiver and the lensing object. With these notations the 
lens equation can be written as \cite{bozza}
\begin{equation}
D_{RS}\tan\beta = \frac{D_{RL}\sin\theta - D_{LS}\sin(\hat{\Theta} - \theta)}{\cos(\hat{\Theta} - \theta)},
\label{eqn.32}
\end{equation}
where $\beta$ denotes the angular position of the source, and $\theta$ denotes 
the angular position of the lensed image of the source as detected by an 
observer. Again, in weak GL, if the source and observer are at infinite 
distances from each other, the above lens equation reduces to \cite{papnoi},
\begin{equation}
\beta = \theta - \frac{D_{LS}}{D_{RS}}\hat{\Theta}.
\label{eqn.32a}
\end{equation}
The angular radius of the Einstein ring is derived by taking $\beta = 0$ 
\cite{papnoi}. Hence, applying this condition to Eq.\ \eqref{eqn.32a}, we obtain the 
Einstein ring radius as
\begin{align}
\theta_E = &\, \frac{D_{LS}}{D_{RS}} \bigg[\frac{4M}{\Upsilon} - \frac{4aM}{\Upsilon^2} + \frac{15\pi M^2}{4\Upsilon^2} - \frac{5\pi aM^2}{\Upsilon^3}\nb\\[5pt]
& + \frac{32M^3}{3\Upsilon^3} + \frac{9\pi a^2 M^2 }{4\Upsilon^4} - \frac{41aM^3}{2\Upsilon^4} + \frac{256a^2 M^3}{5\Upsilon^5} \bigg].
\label{eqn.32b}
\end{align}
\begin{figure}[ht!]
\includegraphics[scale=0.29]{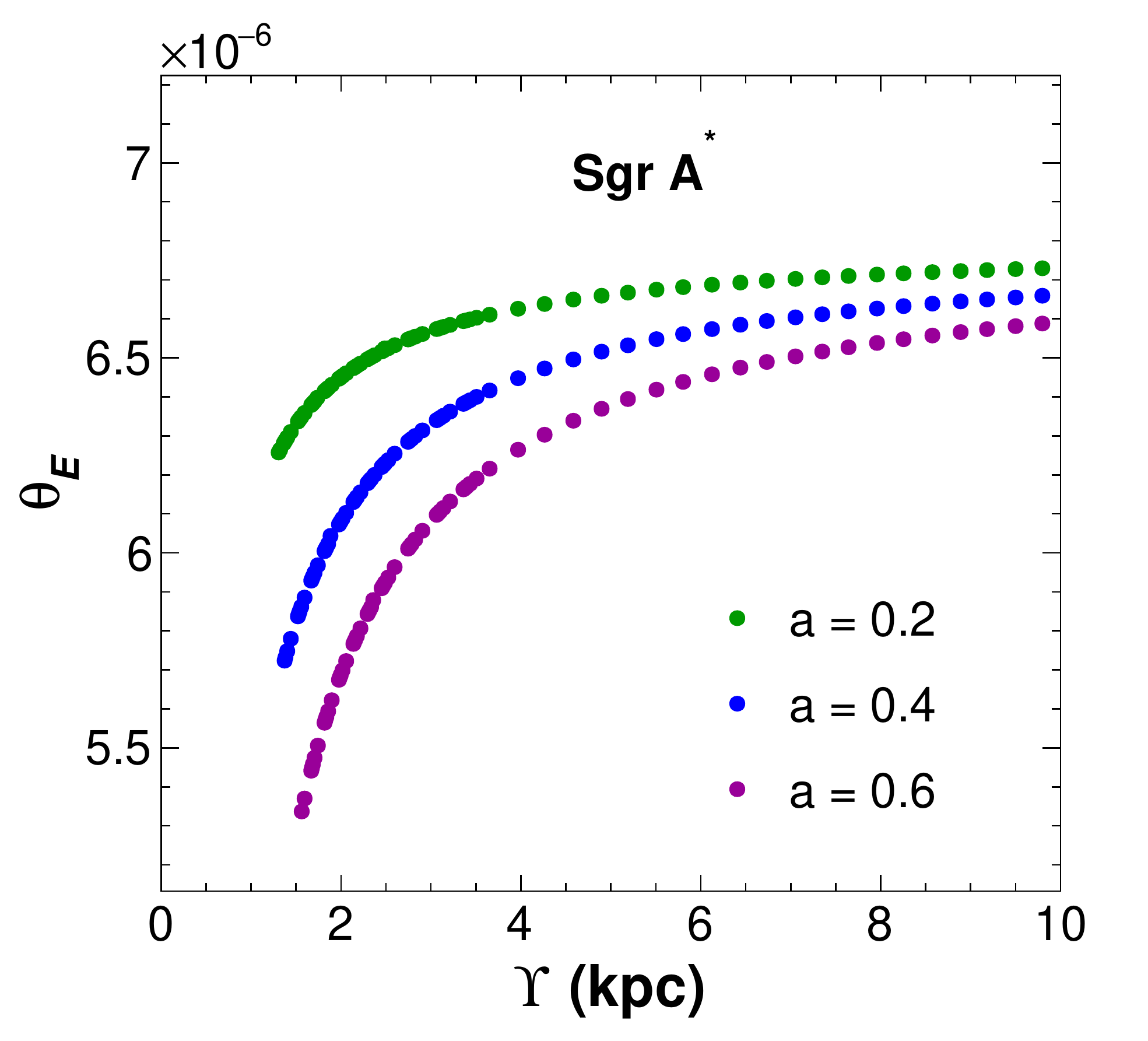}\hspace{1mm}
\includegraphics[scale=0.29]{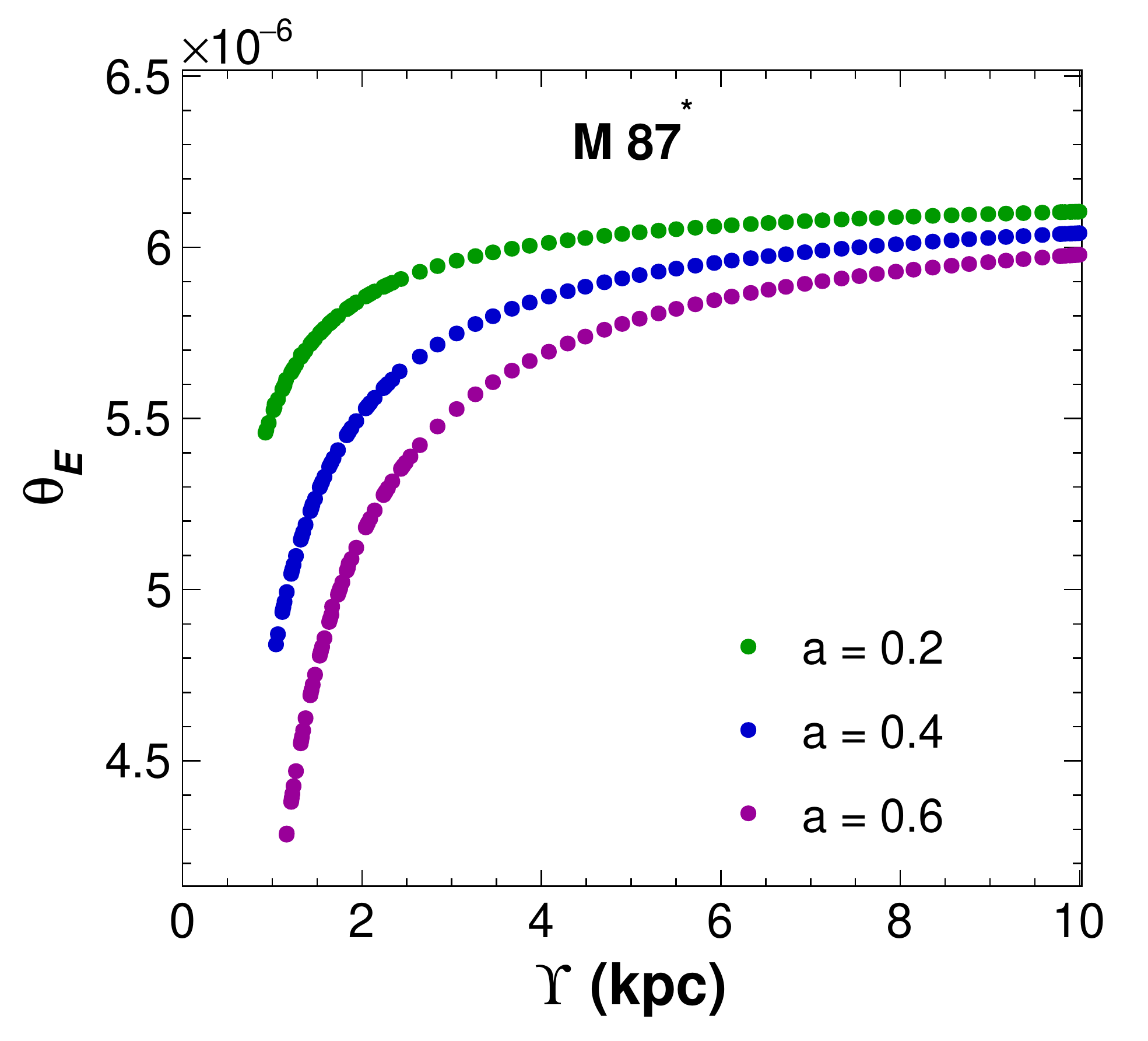}\hspace{1mm}
\includegraphics[scale=0.29]{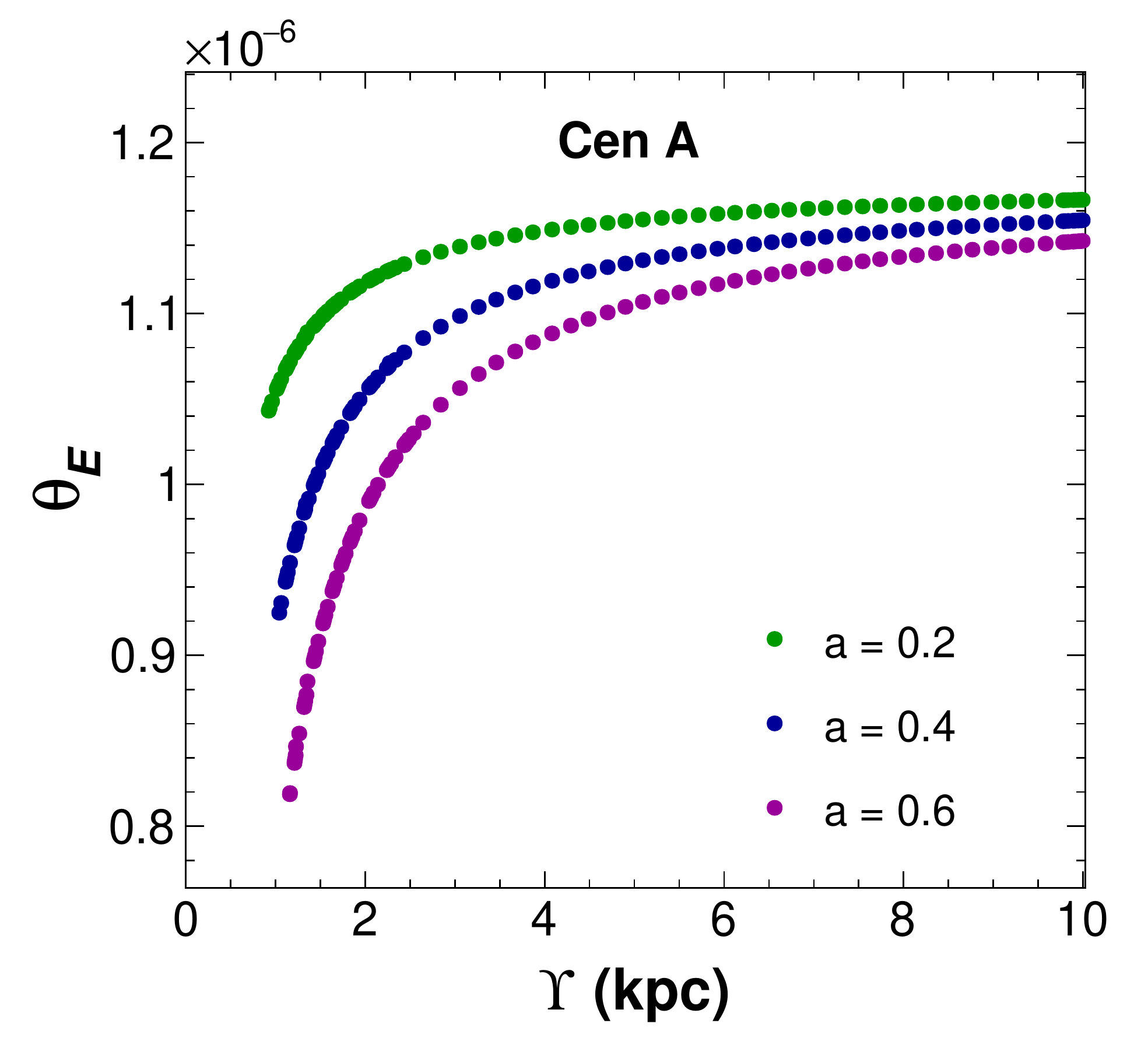}\hspace{1mm}
\caption{Angular radius of the Einstein ring of three black holes, 
SgrA$^*$, M$87^*$ and Cen A as a function of impact parameter for three 
different values of spin parameter, $a = 0.2, 0.4, 0.6$.}
\label{fig.3}
\end{figure}
Furthermore, in weak GL, since the Einstein ring is taken to be small, the 
impact parameter satisfies the relation, $\Upsilon = D_{RL}\sin\theta_E \approx D_{RL} \theta_E$. Hence, we arrive at the final expression of the angular 
radius of the Einstein ring as
\begin{align}
\theta_E =&\, \bigg[ \frac{D_{LS}}{D_{RS}D_{RL}} \bigg(4M - \frac{4aM}{\Upsilon} + \frac{15\pi M^2}{4\Upsilon} - \frac{5\pi aM^2}{\Upsilon^2}\nb \\[5pt]
& + \frac{32M^3}{3\Upsilon^2} + \frac{9\pi a^2 M^2}{4\Upsilon^3} - \frac{41aM^3}{2\Upsilon^3} + \frac{256a^2 M^3}{5\Upsilon^4}\bigg) \bigg]^{1/2}\!\!\!.
\label{eqn.33}
\end{align}

\begin{figure}[ht!]
\includegraphics[scale=0.34]{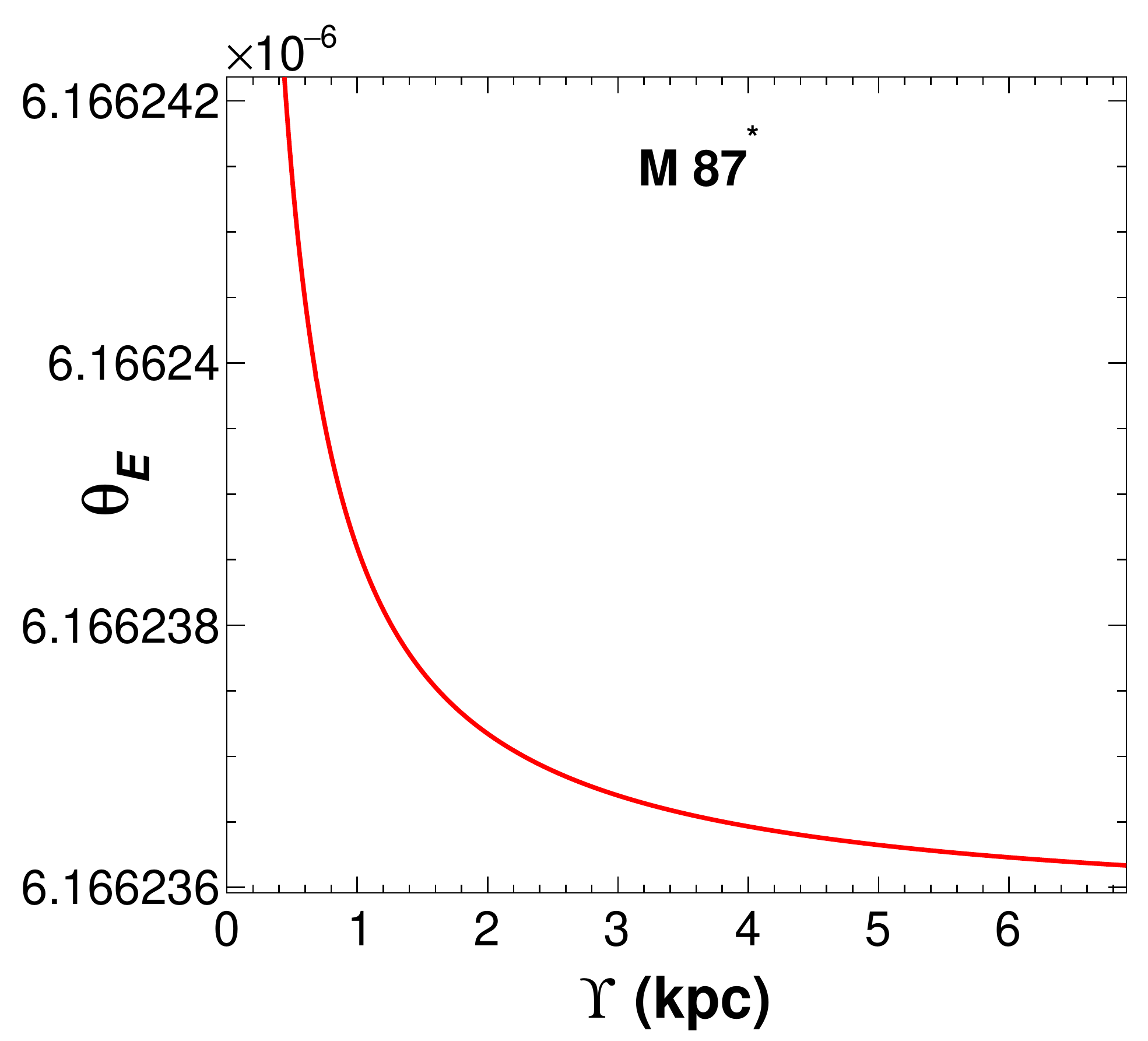}
\caption{Angular radius of the Einstein ring of M $87^*$ black hole as a function of impact parameter for the Schwarzschild case.}
\label{fig.4}
\end{figure}

As in the deflection angle case, we study the effect of the CS gravity on the 
Einstein ring by considering three supermassive black holes, SgrA$^*$, 
M$87^*$ and Cen A as the lens. From Eq.\ \eqref{eqn.33} it is clear that the 
axionic hair of the black hole has some impact on the angular radius of the 
Einstein ring. In Fig.~\ref{fig.3}, the angular radius is depicted as a 
function of the impact parameter for three different values of the spin 
parameter $a = 0.2, 0.4, 0.6$, for each black hole. It is seen that the 
angular radius increases rapidly upto a certain value of the impact parameter 
and then becomes almost constant for higher impact parameter values. 
This behaviour in the angular radius is seen to be similar for all the three 
black holes under consideration. Also, as the spin of the black hole 
increases, the value of the angular radius is seen to decrease. As an
example, the Einstein ring radius of M$87^*$ black hole is depicted for the
Schwarzschild case in Fig.~\ref{fig.4}. It can be seen that the Einstein ring
is found to decrease with the impact parameter for a Schwarzschild black hole,
while for a black hole spacetime in CS gravity, the Einstein ring increases,
thus making it more feasible to be observed in weak lensing techniques.
\begin{figure}[ht!]
\includegraphics[scale=0.29]{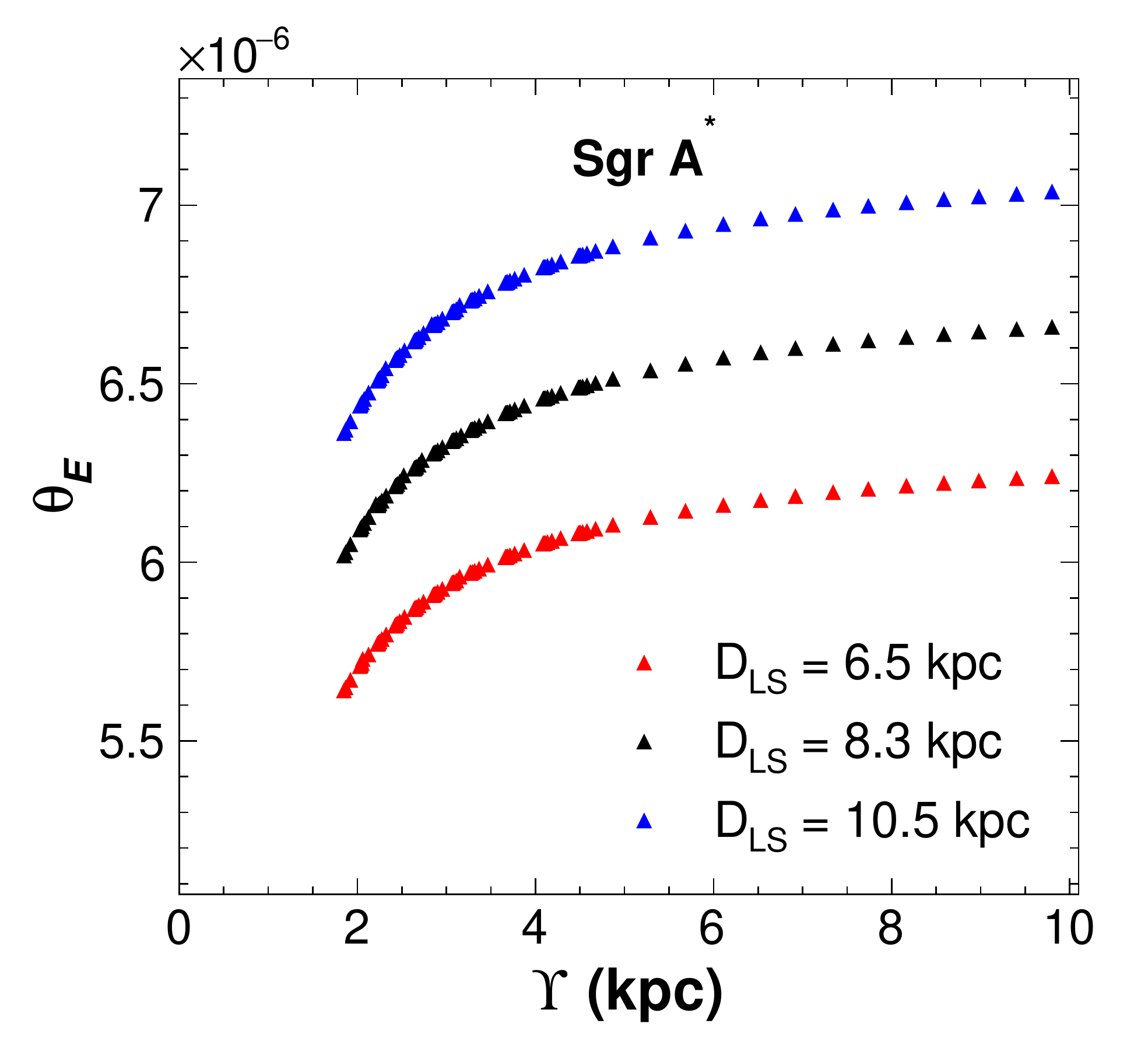}\hspace{1mm}
\includegraphics[scale=0.29]{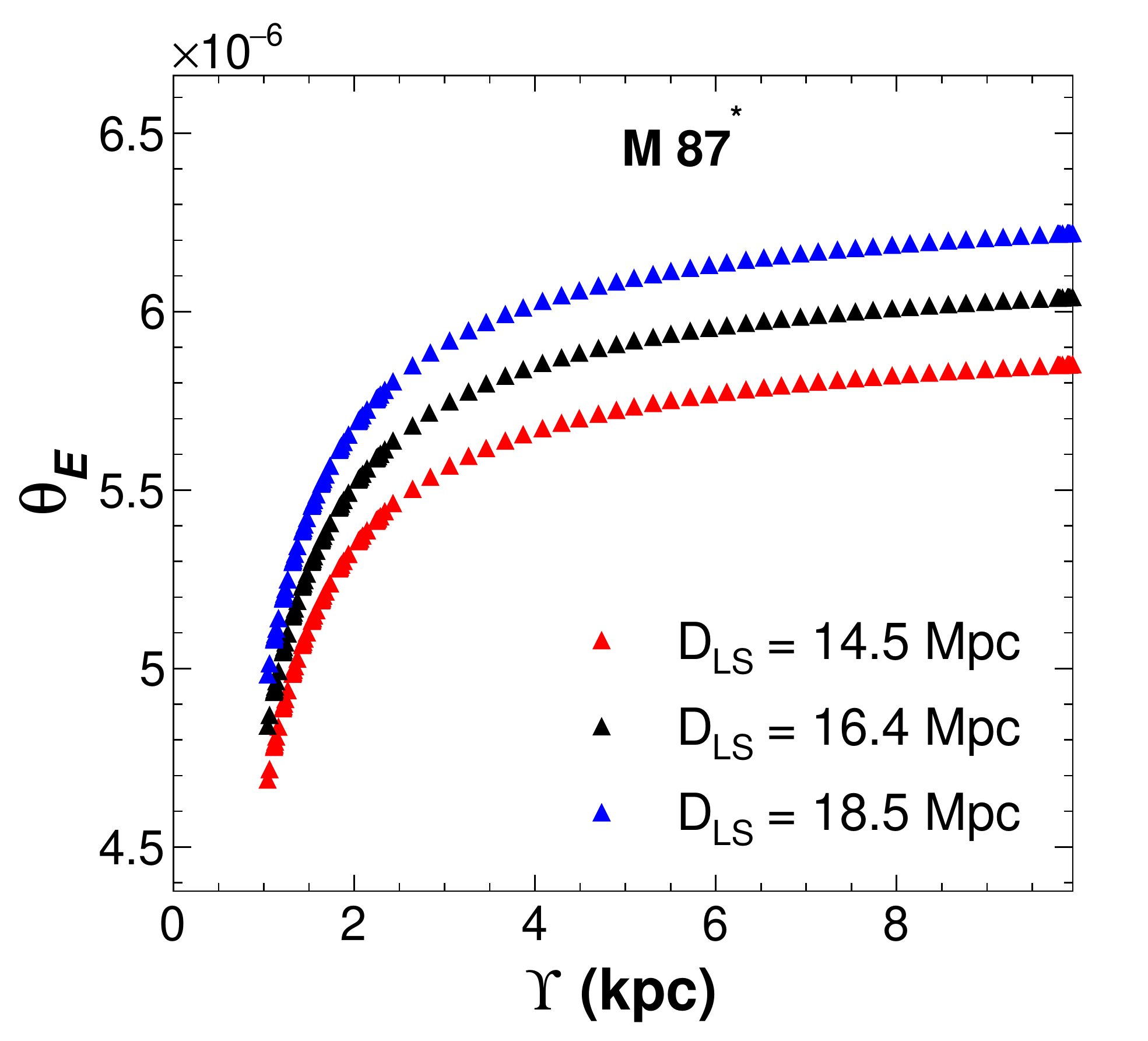}\hspace{1mm}
\includegraphics[scale=0.29]{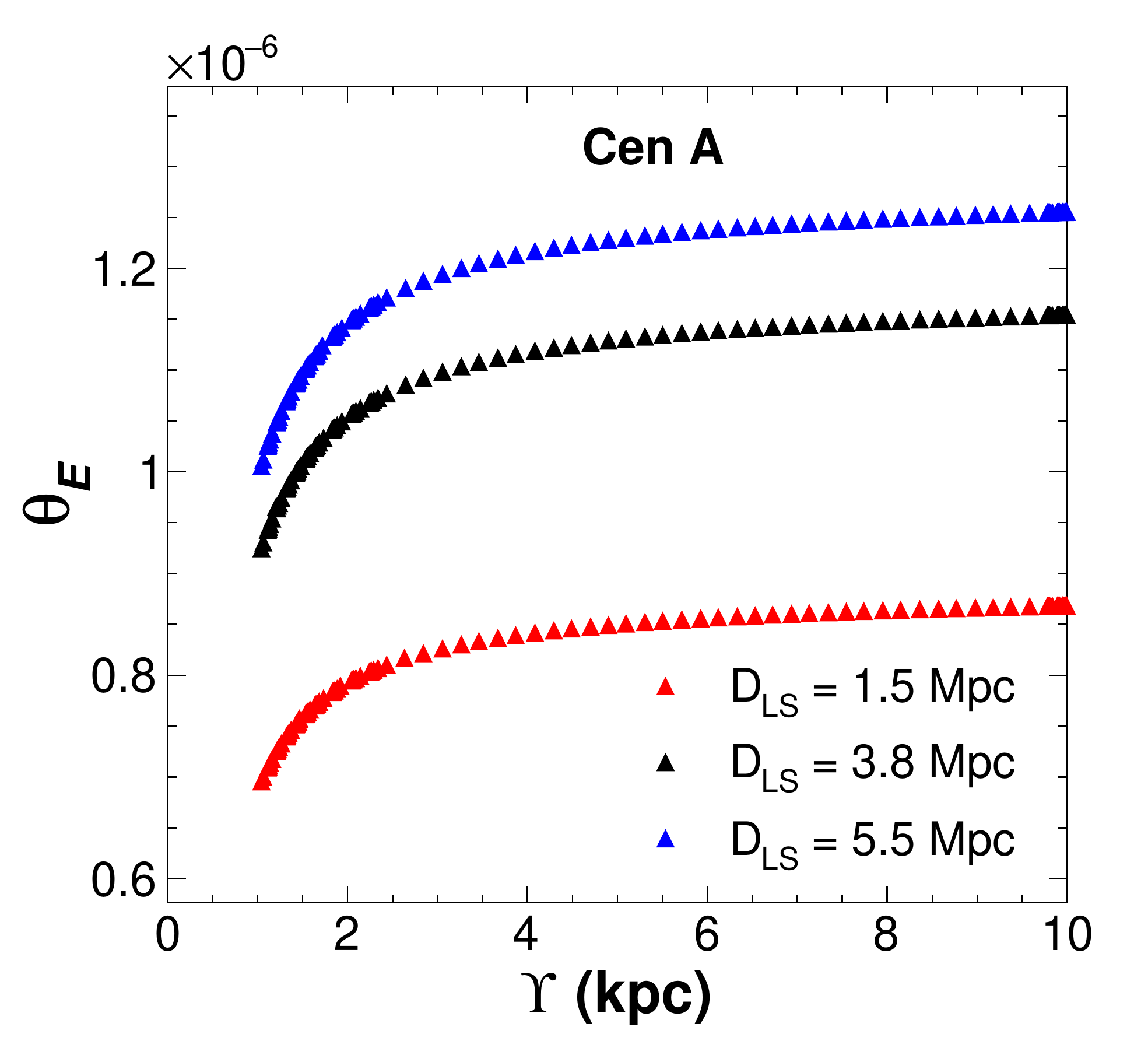}\hspace{1mm}
\caption{Angular radius of the Einstein ring of three black holes, SgrA$^*$, 
M$87^*$ and Cen A as a function of the impact parameter for three different 
values of $D_{LS}$ for each black hole.}
 \label{fig.5}
\end{figure}

Moreover, in Fig.~\ref{fig.5} we study the behaviour of the angular radius of 
the Einstein ring as a function of the impact parameter for three different 
distances of the source to the lens. In the first panel, we have taken the 
SgrA$^*$ black hole, which is located at a distance of $8.3$ kpc from the 
Earth. We consider the source to be situated behind the lens at a 
distance of $6.5, 8.3$ and $10.5$ kpc from the black hole. It can be seen from 
the figure that as the source moves further away, the angular radius becomes 
larger. In the second panel, we consider the M$87^*$ black hole, which is 
located at a distance of $16.4$ Mpc from the Earth. Here, we see that for a 
very small impact parameter, the angular radius for different source distances 
is almost the same. This angular radius becomes higher for the source at far 
away distances as the impact parameter increases. Again, in the third panel, we 
consider the Cen A black hole as a lens, which is situated at a distance of 
$3.8$ Mpc from the Earth. Here also, it can be seen that as the source moves 
far away from the lens, the angular radius becomes higher.
\begin{figure}[ht!]
\includegraphics[scale=0.29]{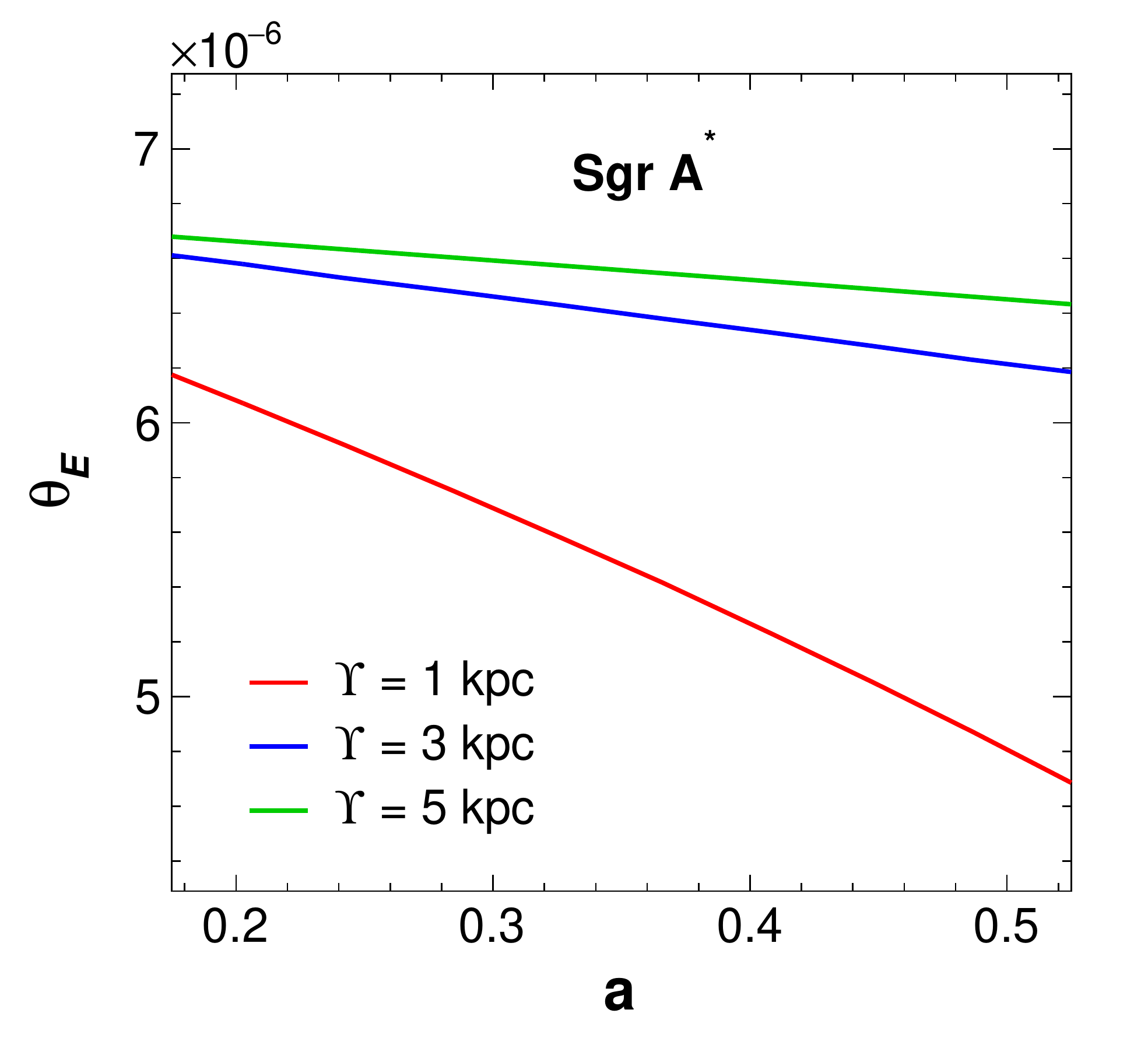}\hspace{1mm}
\includegraphics[scale=0.29]{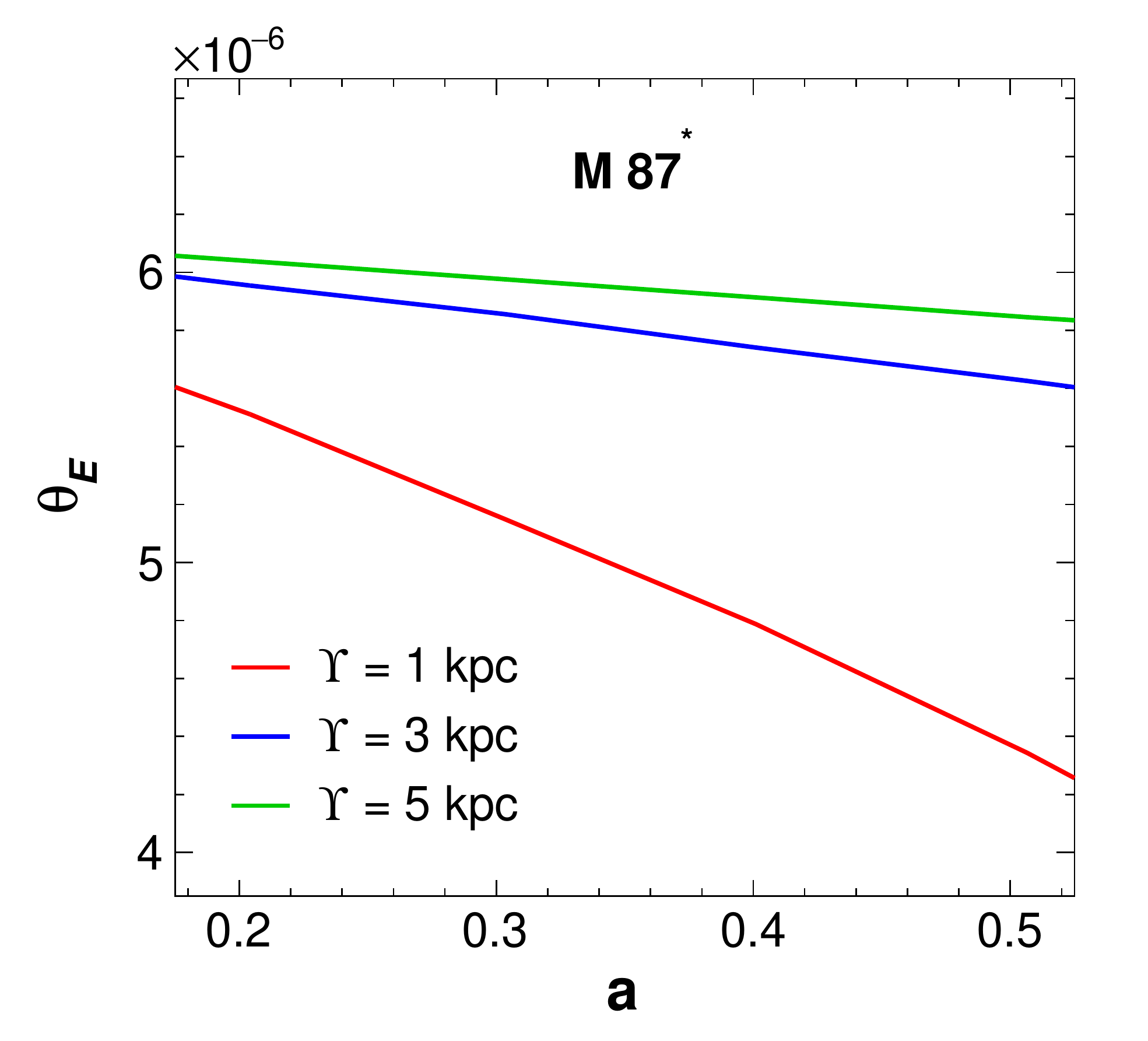}\hspace{1mm}
\includegraphics[scale=0.29]{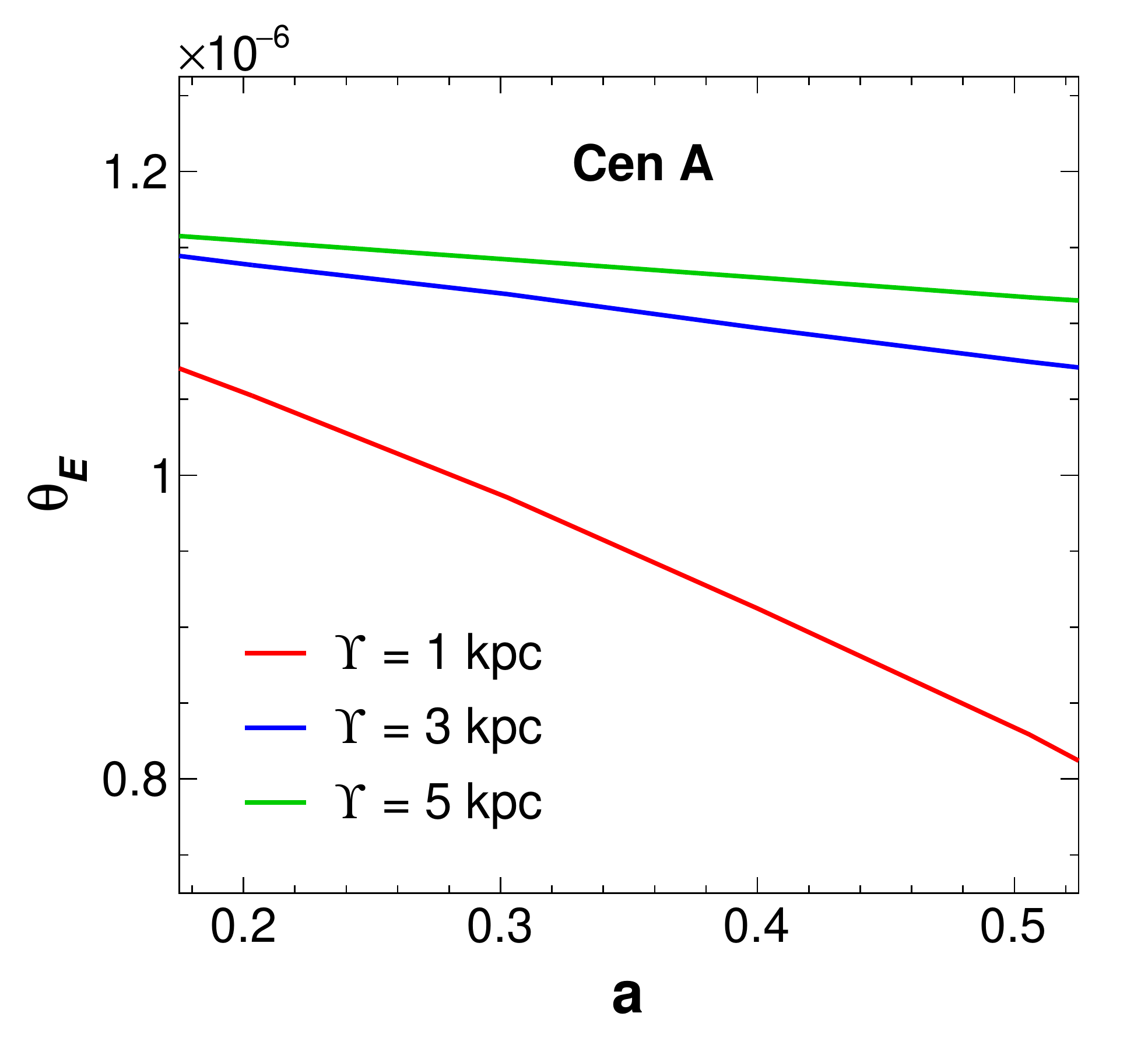}\hspace{1mm}
\caption{Angular radius of the Einstein ring of three black holes, SgrA$^*$, 
M$87^*$ and Cen A as a function of the spin parameter for three different 
values of impact parameter, $\Upsilon = 1, 3, 5$ in units of kpc.}
 \label{fig.6}
\end{figure}

Further, we study the behaviour of the angular radius as a function of the spin 
parameter of the black hole in Fig.~\ref{fig.6}. For each black hole, we take 
three different values of the impact parameter $\Upsilon = 1, 3, 5$ in units 
of kpc. It can be seen that with respect to the spin parameter, the angular 
radius behaves in a similar manner for all three black holes. As the spin 
grows, the angular radius of the Einstein ring decreases. Also, as the value 
of the impact parameter increases, the value of the angular radius becomes 
higher in agreement with the results of Figs.~\ref{fig.3} and \ref{fig.5}.  

\section{Time delay}
\label{sec.5}
If light takes two different paths to travel from a source towards an observer, then it also takes two different times to reach the observer. This difference 
in time to reach the observer is called time delay. In this section, we study 
the effect of the axion coupling parameter on the time delay of light while
they travel from the source to an observer through the neighbourhood of a 
lensed black hole. For this purpose, we rewrite the line element \eqref{eqn.8} 
in the form:
\begin{equation}
ds^2 = - \mathcal{A}(r) dt^2 + B(r) dr^2 + C(r)(d\theta^2 + \sin^2 \theta d\phi^2),
\end{equation}
where
\begin{align}
\mathcal{A}(r) & = \left(1 - \frac{2M}{r}\right) + 2r^2 a \sin^2 \theta W \frac{d\phi}{dt},
\label{eqn.34}\\[5pt]
B(r) & = \left(1 - \frac{2M}{r}\right)^{-1}\!\!,\;\;\; C(r) = r^2.
\label{eqn.34a}
\end{align}
From the expression of impact parameter $\Upsilon$, given by 
Eq.\ \eqref{eqn.20}), we get
\begin{equation}
\frac{d\phi}{dt} = \frac{\left(1 - \frac{2M}{r}\right)\!\Upsilon + r^2 a W}{r^2 - r^2 a W \Upsilon}.
\label{eqn.34b}
\end{equation}
Following Refs.\ \cite{weinberg, jamil, keeton}, the time delay for the slowly 
rotating black hole in Chern-Simons gravity is obtained as
\begin{align}
\Delta T = &\; 2\left(\sqrt{D_{RL}^2 - r_{ps}^2} + \sqrt{D_{LS}^2 - r_{ps}^2}\right) + 4M \log \frac{\left(D_{RL} + \sqrt{D_{RL}^2 - r_{ps}^2}\right) 
\left(D_{LS} + \sqrt{D_{LS}^2 - r_{ps}^2}\right)}{r_{ps}^2}\nb\\[5pt] 
& - \frac{4 a \Upsilon M}{r_{ps}^2} \left\lbrace \left(3 + 2\,\frac{r_{ps}}{D_{RL}}\right) \sqrt{\frac{D_{RL} - r_{ps}}{D_{RL} + r_{ps}}} + \left(3 + 2\,\frac{r_{ps}}{D_{LS}}\right) \sqrt{\frac{D_{LS} - r_{ps}}{D_{LS} + r_{ps}}}\, \right\rbrace.
\label{eqn.35}
\end{align}

\begin{figure}[ht!]
\includegraphics[scale=0.365]{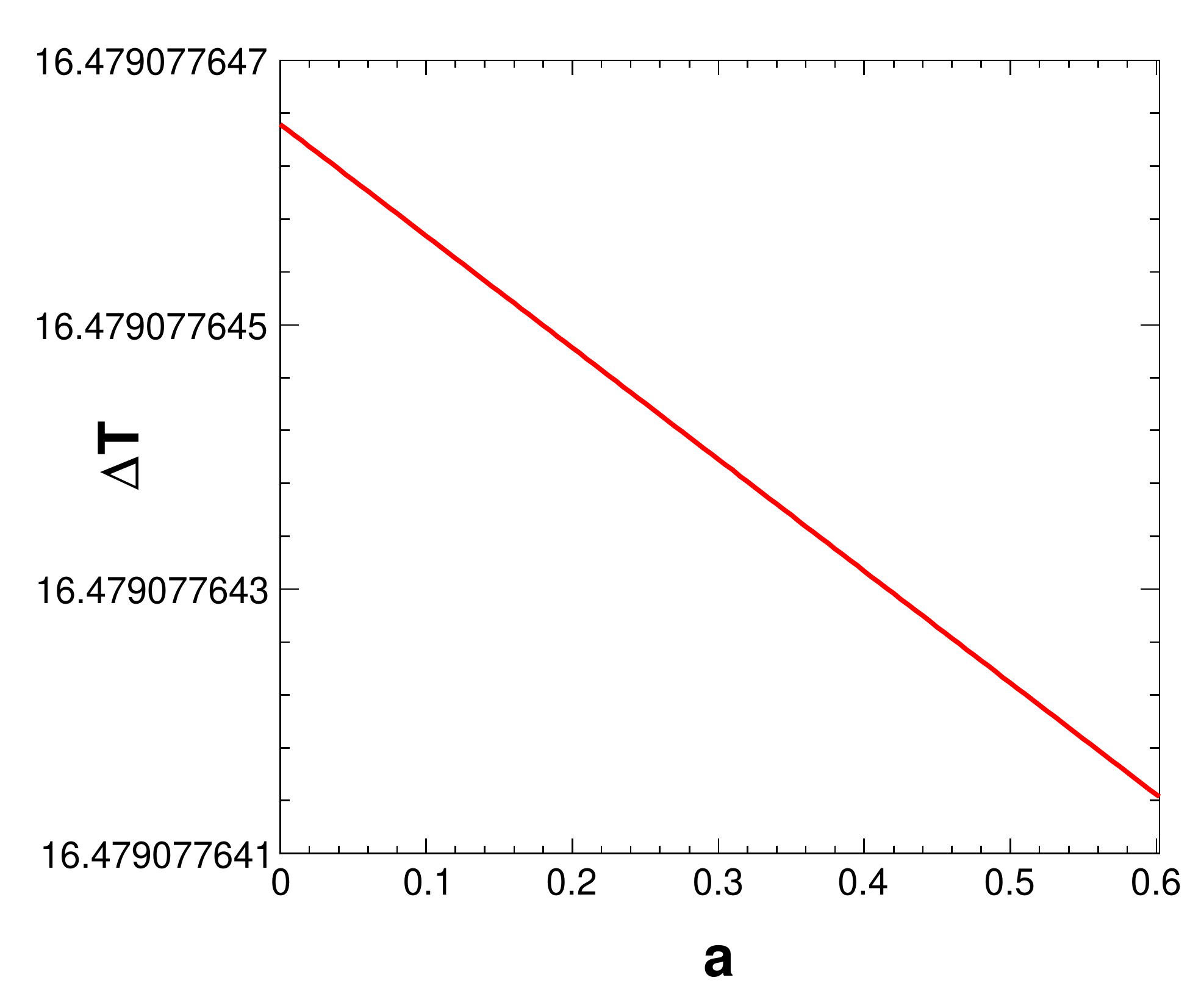}\hspace{3mm}
\includegraphics[scale=0.34]{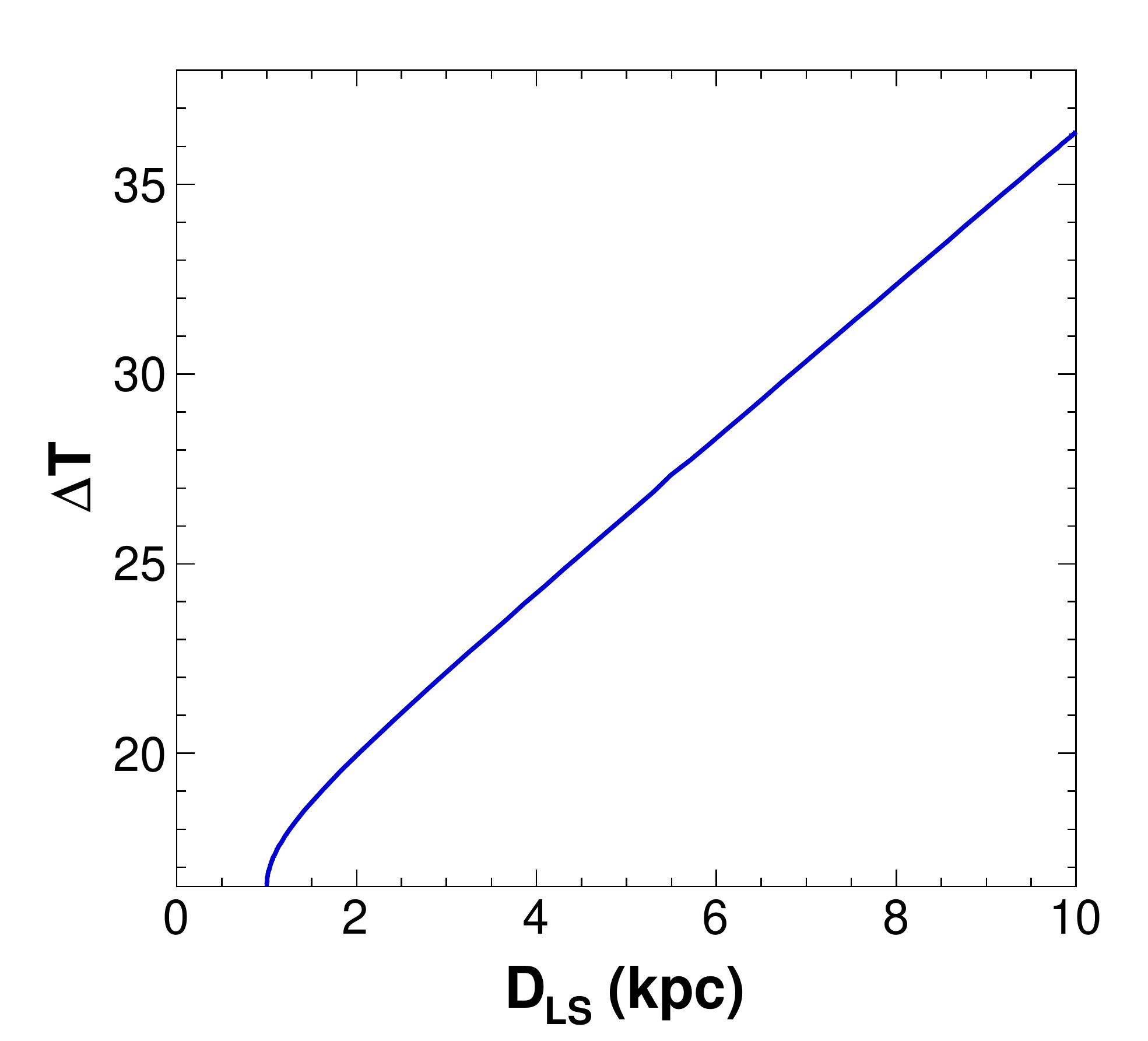}\hspace{1mm}
\caption{Time delay $\Delta T$ as a function of spin parameter $a$ in the left 
panel and as a function of source distance $D_{LS}$ in the right panel for 
the SgrA$^*$ black hole.}
 \label{fig.7}
\end{figure}

\begin{table}[ht!]
\begin{center}
\caption{Estimated time delay for supermassive black holes in Chern-Simons 
gravity.}
\vspace{0.2cm}
\begin{tabular}{ c c c c c }
\hline \hline 
Galaxy & Distance & Mass & $\Delta T$ & $\Delta T$ (Schwarzschild case)\\ \vspace{0.5mm}
  		& (Mpc) & ($M_\odot$) & (hours) & (hours)\\ 
   \hline 
SgrA$^*$ 	&	$0.0083$	&	$4 \times 10^6$		&	0.27 & 0.19\\
M87$^*$		&	$16.4$	&	$6.5 \times 10^9$	&	546.67 & 274.05\\
Cen A	&	$3.8$	&	$5.5 \times 10^7$	&	126.67 & 2.53\\
 \hline \hline   
\end{tabular}
\label{Table:1}
\end{center}
\end{table}

It is clear from Eq.\ \eqref{eqn.35} that the axionic hair of the black hole 
has some effect on the time delay. Fig.~\ref{fig.7} shows the behaviour of the 
time delay of light with respect to the spin parameter $a$ and the source 
distance $D_{LS}$ for the Sgr A$^*$ black hole. It can be seen in the left 
panel that with the growth of the spin of the black hole, the time delay 
decreases. On the other hand, in the right panel, it is seen that when the 
source is assumed to move further away from the lens, the time delay 
increases. Decreasing time delay with spin of the black hole has also been 
studied in \cite{mancini} in GR. The estimated time delay for three 
supermassive black holes have been displayed in Table~\ref{Table:1}. One can 
see that being the most massive and farthest among the considered black holes, 
M87$^*$ produced the maximum time delay amongst the black holes.

\section{Optical behaviour of the black hole: Shadow}
\label{sec.6}
The black hole shadow, a dark area that occupies the center of the bright 
accretion disk, is an impressive visual characteristic of a black hole. The 
shadow is a consequence of the black hole's intense gravitational field, which 
deflects and confines light, ultimately creating a photon sphere \cite{adair, Gammon, ronit23, parbin23, papnoi}. The size and 
shape of the shadow of a black hole depend on its mass, rotation and 
proximity to Earth, presenting an exceptional opportunity to scrutinize the 
features of black holes. The utilization of the Event Horizon Telescope to 
investigate the black hole shadow has furnished compelling evidence of black 
hole existence and tested the accuracy of GR in the powerful gravitational 
domain \cite{akiyama, EHT02}. The study of the black hole shadow as an optical attribute of black 
holes is a swiftly advancing domain that holds the potential to enhance our 
comprehension of gravity and the fundamental nature of spacetime.

\subsection{Null geodesics and photon sphere}\label{sec:Geodesics}
To explore the shadow of the black hole defined by the metric \eqref{eqn.8} in 
CS gravity, first we start with the investigation of geodesics in the theory. 
To this end, it is most expedient to employ the Lagrangian framework. The 
Lagrangian associated with the theory is expressed as
\be
L=\frac{1}{2}g_{\mu\nu}\dot{x}^{\mu}\dot{x}^{\nu},
\ee
where $\dot{x}^{\mu}$ denotes the derivative of the coordinate $x^{\mu}$ 
with respect to the affine parameter $\tau$. Utilizing this Lagrangian, we can 
define the conjugate momenta and the Hamiltonian of the system as
\be
p_{\mu}=g_{\mu\nu}\dot{x}^{\nu}\, ,\quad H=\frac{1}{2}g^{\mu\nu}p_{\mu}p_{\nu}\, .
\ee
The Hamilton-Jacobi equation is given by
\be
\frac{\partial S}{\partial \tau}=\frac{1}{2}g^{\mu\nu}\frac{\partial S}{\partial x^{\mu}}\frac{\partial S}{\partial x^{\nu}}\, .
\ee
One may note that the Hamiltonian is independent of the variables $t$, 
$\phi$ and $\tau$ explicitly, which gives us the privilege to write the 
expression for the action as \cite{adair}
\be
S=-\frac{1}{2}\xi^2\tau - \mathcal{E} t+L_z \phi+\tilde S(r,\theta)\, ,
\ee
where the terms $L_z$, $\xi^2$ and $ \mathcal{E}$ are constants. 
Now, as earlier for a slowly rotating black hole, we assume that the spin
parameter $a$ of the black hole is comparatively smaller and we consider
that $\tilde S(r,\theta)=\tilde S_{r}(r)+\tilde S_{\theta}(\theta)$. 
Implementing these assumptions in the Hamilton-Jacobi equation, we get
\be
\left(\frac{\partial S_{\theta}}{\partial\theta}\right)^{\!2}+\frac{L_z^2}{\sin^2\theta}=-r (r-2 M) \left(\frac{\partial S_{r}}{\partial r}\right)^{\!2} -\frac{ \mathcal{E}^2 r^3}{2 M-r}  -\frac{2 a r^3 W(r) \mathcal{E}L_z}{2 M-r}- r^2\xi^2\, .
\ee
In deriving this expression we neglect the higher-order terms of $a$ as they
have negligible contributions being $a$ is a small quantity. 

As the left-hand side of the above equation solely relies on $\theta$ and the 
right-hand side only on $r$, it can be inferred that both sides of the 
equation are equivalent to a constant value, say $j^2$. Using this condition, 
all the derivatives of $S$ can be expressed as
\begin{eqnarray}
\frac{\partial S}{\partial r}&=&\pm\sqrt{\frac{2 a  \mathcal{E} r^3 L_z W(r)+r^2 \big( \mathcal{E}^2 r+\xi ^2 (2 M-r)\big)+j^2 (2 M-r)}{r (r-2 M)^2}}\, ,\\
\frac{\partial S}{\partial \theta}&=&\pm\sqrt{j^2-\frac{L_z^2}{\sin^2\theta}}\, ,\\
\frac{\partial S}{\partial \phi}&=&L_z\, ,\\
\frac{\partial S}{\partial t}&=&- \mathcal{E}\, .
\end{eqnarray}
These derivatives of $S$ can be further used to obtain the orbit equations 
associated with the black hole spacetime. We use  
$\partial_{\mu}S\rightarrow p_{\mu}$, where four-momenta $ p_{\mu}$ has the 
following explicit expressions for the black hole considered in this study:
\begin{align}
p_{r}&=\frac{\dot{r}}{1-2M/r}\, , \\  
p_{\theta}&=r^2\dot{\theta}\, , \\
p_{\phi}&=a r^2 W(r)\sin^2\theta \dot{t}+r^2\sin^2\theta \dot{\phi}\, , \\
p_{t}&=-(1-2M/r)\dot{t}+a r^2 W(r)\sin^2\theta \dot{\phi}.
\end{align}
Using these equations, we can further obtain a set of first-order differential 
equations in terms of the conserved quantities $ \mathcal{E}, L_z, \xi$ and 
$j$ as
\begin{eqnarray}\label{geodesicEqns}
\dot{r}^2&=&-(1-2M/r)\left(\xi^2+\frac{j^2}{r^2}\right)+ \mathcal{E}^2+2 a  \mathcal{E} L_z W(r)\, ,\\
r^2\dot{\theta}&=&\pm\sqrt{j^2-\frac{L_z^2}{\sin^2\theta}}\, ,\\ \label{rEq}
r^2\dot{\phi}&=&\frac{L_z}{\sin^2\theta}+\frac{a  \mathcal{E} r^3 W(r)}{2 M-r}\, ,\\
r^2\dot{t}&=& -\frac{r^3 \left(a L_z W(r)+ \mathcal{E}\right)}{2 M-r}\, ,
\end{eqnarray}
These equations are referred to as the equations of geodesics. It is evident 
that in the asymptotic limit, $j^2$ corresponds to the overall angular 
momentum of the orbit, while $L_z$ denotes the angular momentum component 
along the $z$ axis. 

The null geodesics refer to paths followed by massless particles (such as 
photons) in curved spacetime. To describe these paths, we typically use a set 
of coordinates and an affine parameter $\tau$ that represents the 
\textit{affine distance} travelled along the path. The null geodesics satisfy 
a specific condition, $p_\mu p^\mu \equiv \xi^2 = 0.$ 
One can rescale the affine parameter $\tau$ in such a way that the energy 
$ \mathcal{E}$ of the photon is equal to one. Using this information, we can 
write an equation for the radial coordinate of the photon's path, which takes 
the form \cite{15s,16s,17s,18s,102-9, ronit23, parbin23}:
\begin{equation}
\dot{r}^2+V_{ ph}(r)=0 ,
\end{equation}
where $V_{  ph}(r)$ is a function of the radial coordinate $r$. Specifically, 
for our present case this function takes the form:
\begin{equation}
V_{  ph}(r)=-2 a L_z W(r)+\frac{j^2 (r-2 M)}{r^3}-1.
\end{equation}

The photon sphere is a region of space that is defined by constant-$r$ photon 
orbits, i.e.\ photons following these paths have a fixed value of $r$. To find 
the radius of the photon sphere, we need to find the value of $r$ that 
satisfies two conditions: $V_{  ph}(r_{  ps})=0$ and $V'_{  ph}(r_{  ps})=0$, 
where the prime denotes a derivative with respect to $r$ \cite{ronit23, parbin23}. These two conditions 
give us the radius of the photon sphere $r_{  ps}$ as well as the value of 
$j_{  ps}^2$, which is a constant related to the angular momentum of the 
photon. In this context, it is to be mentioned that up to the linear order of 
$a$, one can have the following relations:
\begin{equation}
r_{  ps}=r_{  ps}^{(0)}+a r_{  ps}^{(1)}\, ,\quad j_{  ps}^2=(j_{  ps}^{(0)})^2+a(j_{  ps}^{(1)})^2, 
\end{equation}
where $r_{  ps}^{(0)}$ is defined by the equation \cite{adair},
\begin{equation}
r_{  ps}^{(0)}f'(r_{  ps}^{(0)})-2 f(r_{  ps}^{(0)})=0\, ,
\end{equation}
when $a=0$.
Using these relations, one can find the following explicit expressions of
$r_{  ps}$ and $j_{  ps}^2$ for the black hole considered in this study:
\begin{eqnarray}
r_{ps}&=&3 M+\frac{2 a L_z \left(81 M^4-31 \eta \kappa ^2\right)}{729 M^5} \, ,\\[5pt]
j_{  ps}^2&=&27 M^2+a L_z \left(4-\frac{131 \eta \kappa ^2}{189 M^4}\right) . \label{jexpression}
\end{eqnarray}
For a static black hole, these two equations give the usual expressions of 
photon radius and angular velocity of photons respectively as 
$r_{ps}^{(0)} = 3M$ and $\omega_{ps}^{(0)} = 1/3 \sqrt{3} M$.

\subsection{Black hole shadow}\label{sec:shadow}
Here for simplicity and convenience, we suppose there is an observer located 
at a distance $r_{obs}$ from a black hole and at a polar angle $\theta_{0}$ 
with $\phi_{0}=0$. To determine the path of a photon moving towards the 
observer in the direction $dr/dt>0$, we can use the angular momentum 
parameters $j^2$ and $L_z$. Our objective is to find the angle at which the 
photon hits the observer's plane perpendicular to the $r$-direction. To do so, 
we need the tangential vector at that specific point in space, which can be 
expressed as \cite{adair, Gammon}
\begin{equation}
\mathbf{U}=-\,\dot{r}\,\mathbf{e}_{r}+r_{obs}\dot{\theta}\,\mathbf{e}_{\theta}+r_{obs}\sin\theta_0\dot{\phi}\,\mathbf{e}_{\phi}\, ,
\end{equation}
where we introduce an orthonormal system for the observer. This system 
consists of three unit vectors:
$$\mathbf{e}_{r}=-\,\partial_{r}\,,\;\; 
\mathbf{e}_{\theta}=\frac{\partial_{\theta}}{r_{obs}}\,,\;\; 
\mathbf{e}_{\phi}=\frac{\partial_{\phi}}{r_{obs}\sin\theta_0} \, .$$ 
Moreover, we assume that the observer is looking directly towards the black 
hole. Let's define the angle of incidence of the photon on the plane 
$r=r_{obs}$ as $\pi/2-\delta$ and the angle that the projected vector forms 
with the direction $\mathbf{e}_{\phi}$ as $\alpha$. In other words, we can 
express the 
tangent vector as $$\mathbf{U}=-\,\dot{r}\,\mathbf{e}_{r}+\sin\delta\left(\mathbf{e}_{\theta}\sin\alpha + \mathbf{e}_{\phi}\cos\alpha\right).$$ To obtain the 
values of $\sin\delta$ and $\cos\alpha$, we introduce the following 
parametrization \cite{adair, Gammon}: $$\sin\delta=r_{obs}\sqrt{\dot{\theta}^2+\sin^2\theta_0\dot{\phi}^2}\;\;\; \text{and}\;\;\; \cos\alpha=\frac{\sin\theta_0\dot{\phi}}{\sqrt{\dot{\theta}^2+\sin^2\theta_0\dot{\phi}^2}}\,.$$
Further, we assume that $r_{obs}\sqrt{\dot{\theta}^2+\sin^2\theta_0\dot{\phi}^2}\ll1$ in these expressions because we are considering the limit where 
$r_{obs}$ approaches infinity. By using the geodesic equations, we can relate 
these angles to the angular momentum, or alternatively, can express the 
angular momentum of the geodesic in terms of these angles, given by \cite{adair, Gammon}
\begin{equation}
j=r_{obs}\sin\delta\, ,\quad L_z=r_{obs}\sin\theta_0\cos\alpha\sin\delta.
\end{equation}
From the above results and using Eq.\ \eqref{jexpression}, one can obtain 
the relation,
\begin{equation}
r_{obs}^2\sin^2\delta=27 M^2+a \sin \theta_0  \left(4-\frac{131 \eta \kappa ^2}{189 M^4}\right)\cos\alpha\;  r_{obs}\sin\delta\,.
\end{equation}
This equation is responsible for determining the shape of the black hole shadow
contour, denoted by $\delta(\alpha)$. Upon performing a linear expansion in 
$a$, the solution to this equation can be expressed as follows:
\begin{equation}\label{shadoweqn}
r_{obs}\sin\delta=3 \sqrt{3} M+a \sin \theta_0 \left(2-\frac{131 \eta \kappa ^2}{378 M^4}\right)\cos\alpha\,.
\end{equation}
This equation can be used to find the value of $\delta$ as a function of 
$\alpha$, which then characterizes the shape of the shadow. One may note that 
the curve $\delta(\alpha)$ is an approximation to the circumference of radius 
$R_{sh}=3 \sqrt{3} M$ centered at $\alpha=0$, with an assumption that  
$\delta <<1$. 

\begin{figure}[!ht]
   \centerline{
   \includegraphics[scale = 0.65]{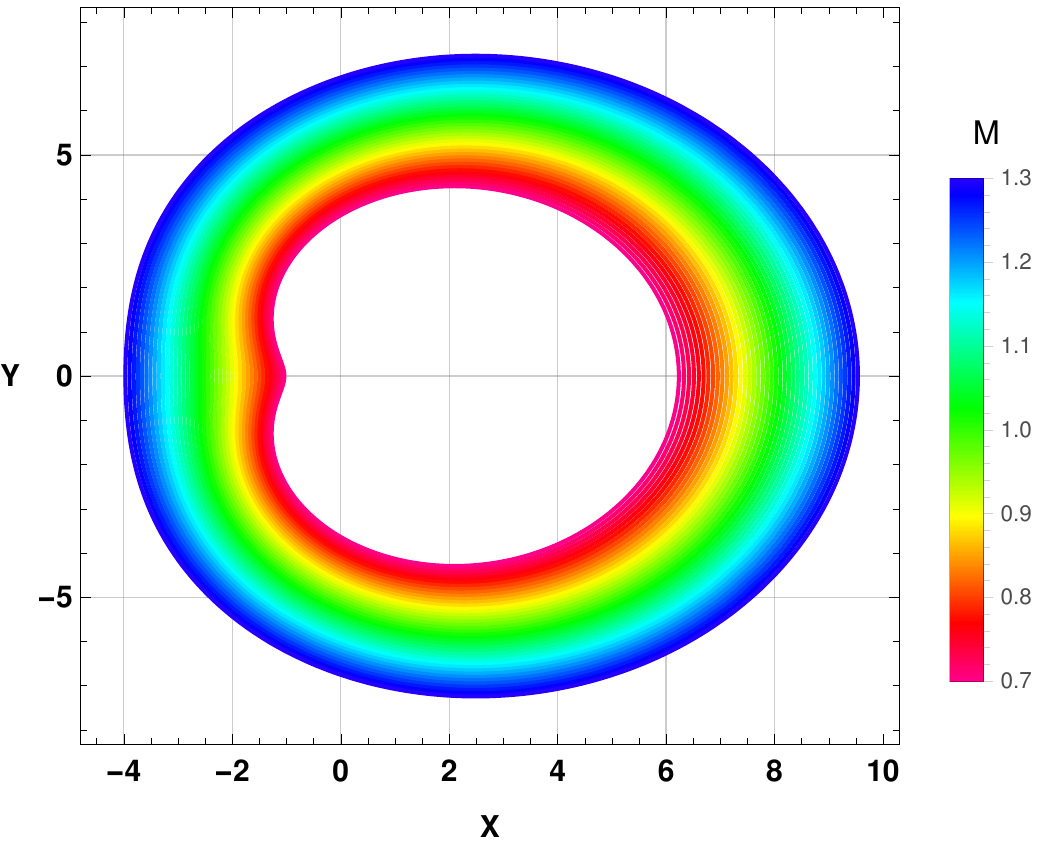}\hspace{0.5cm}
   \includegraphics[scale = 0.65]{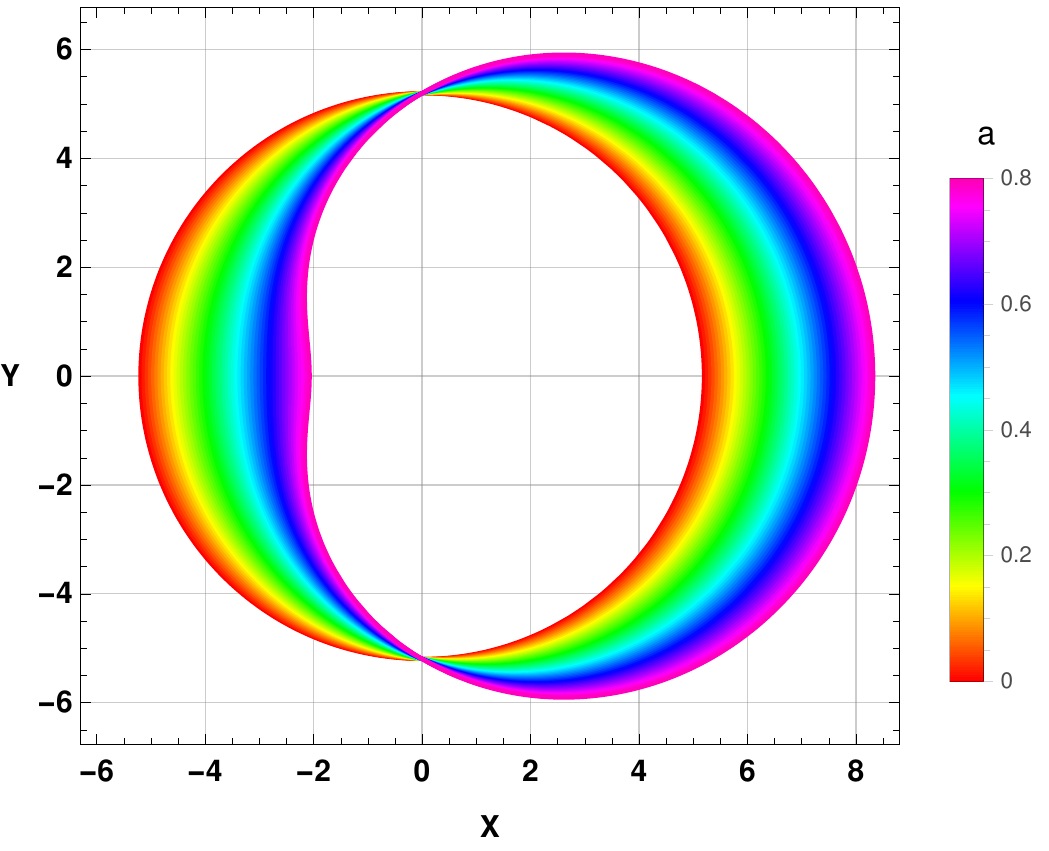}}
\vspace{-0.2cm}
\caption{Stereographic projection of the black hole shadow. On the left panel, 
we consider $a=0.7, \theta = \pi/2$ and $ \eta = 0.1$ for different values
of the black hole mass $M$ and on the right panel, $M=1, \theta = \pi/2$ and 
$ \eta = 0.1$ for different values of spin parameter $a$. In both cases, we 
use $\kappa = 1.$ }
\label{fig_shadow01}
\end{figure}

\begin{figure}[!ht]
   \centerline{
   \includegraphics[scale = 0.65]{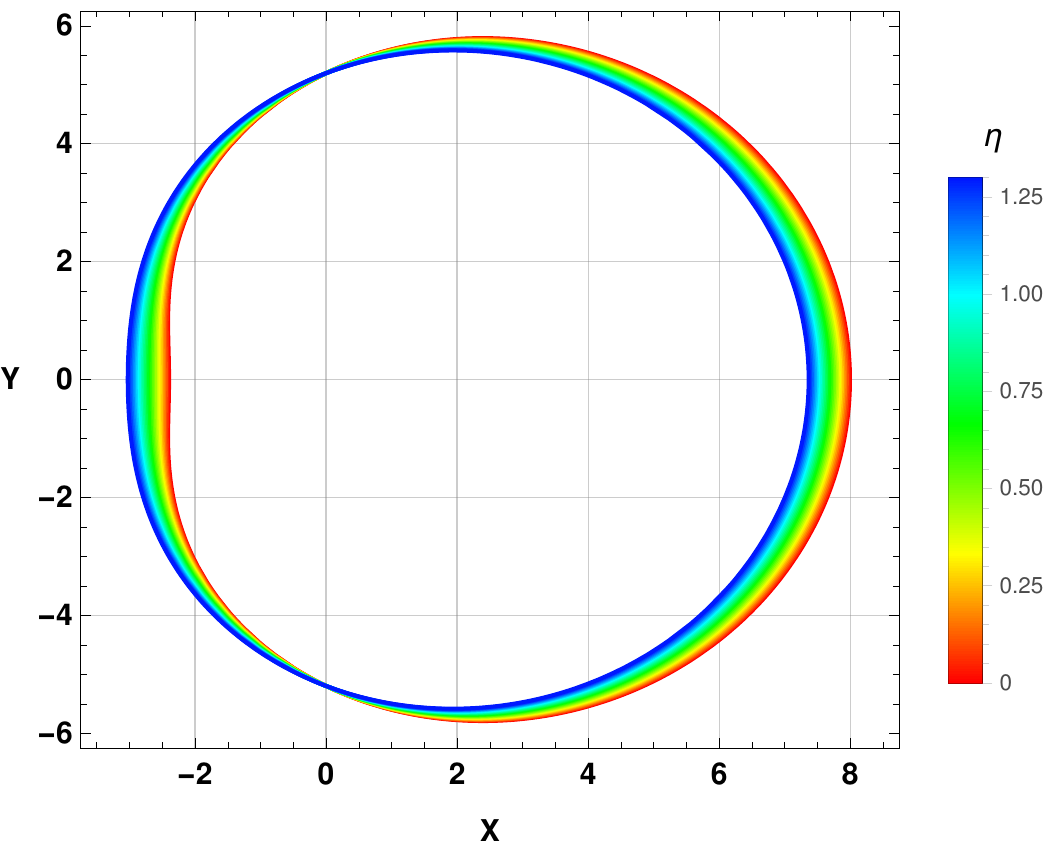}\hspace{0.5cm}
   \includegraphics[scale = 0.65]{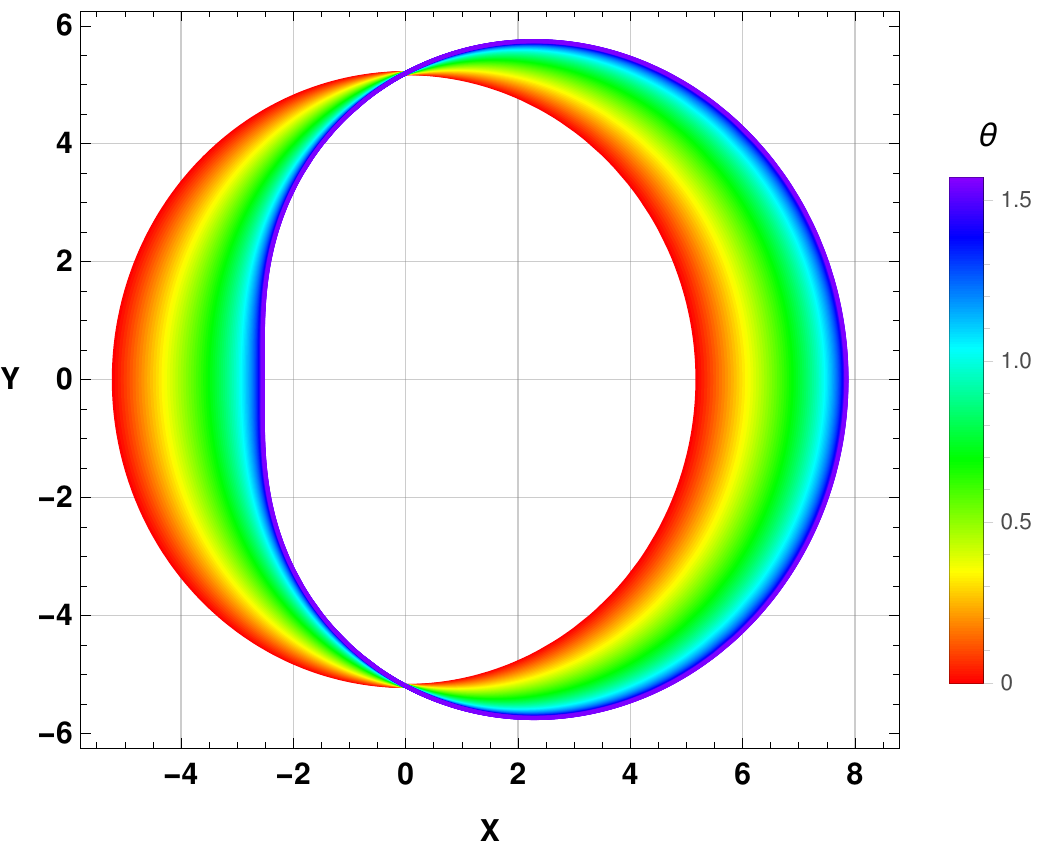}}
\vspace{-0.2cm}
\caption{Stereographic projection of the black hole shadow. On the left panel, 
we consider $a=0.7, \theta = \pi/2$ and $ M = 1$ for different values of CS 
coupling parameter $\eta$ and on the right panel, $M=1, \eta = 0.3$ and 
$ a = 0.7$ for different values of $\theta$. In both cases, we use 
$\kappa = 1.$ }
\label{fig_shadow02}
\end{figure}

We use Eq.\ \eqref{shadoweqn} to obtain the stereographic projections of the 
black hole shadow in Figs.\ \ref{fig_shadow01} and \ref{fig_shadow02}. On the 
left panel of Fig.\ \ref{fig_shadow01}, we show the black hole shadow for 
different values of black hole mass $M$. It is clear from the figure that with 
an increase in the value of $M$, the shadow radius also increases gradually in 
accordance with that of the standard static Schwarzschild black hole. On the 
right panel of Fig. \ref{fig_shadow01}, we can see that with an increase in 
the black hole spin parameter $a$, the shadow of the black hole gets deformed 
as expected. Although in the figure, we consider comparatively higher values 
of the parameter $a$ for the graphical representation purposes, in the case of 
a slowly rotating black hole, $a$ is a small quantity and hence a minimal 
distortion in the shadow is expected. Thus, from the shadow behaviour, one can 
estimate the spin parameter. Finally, in Fig.\ \ref{fig_shadow02}, we show
the impacts of the axionic coupling parameter $\eta$ on the shadow of the black
hole along with the variation in the shadow with respect to $\theta$. One can
see that $\eta$ affects the shadow minimally, and with an increase in the
value of $\eta$, the deformation of the shadow decreases very slowly. It
implies that the parameter $\eta$ may counter the effects of the black hole
spin $a$. It is seen that $\theta$ also has significant impacts on the
appearance of the shadow as expected.

\section{Summary and outlook}
\label{sec.7}
One of the popular extensions of GR is the CS modified gravity. The scalar 
field, in this case the axion field, is a basic feature of this widely known 
modified theory of gravity. In this paper, we study the effect of this axion 
field on the weak gravitational lensing. For this purpose, we first consider 
an exact slowly rotating Kerr-type black hole spacetime with an axionic hair. 
We then obtain the weak lensing angle of light around the black hole. Next, we 
study the behaviour of the Einstein ring formed due to deflection of light by 
the Kerr-type black hole. Then we move forward to obtain the time delay of 
light due to the presence of the black hole in its path. Finally, we study the 
impact of the axion coupling parameter on the shadow cast by the black hole 
under consideration.

To obtain the deflection angle of light around the black hole, we implement 
the extension of Gibbons-Werner method which was studied by Ishihara \etal in 
2016. This method is independent of asymptotic flatness of the spacetime. We 
study the behaviour of the deflection angle as a function of the impact 
parameter for a set of three black holes, viz.\ SgrA$^*$, M$87^*$ and Cen A. 
We compare the deflection angle for the Kerr-type black hole in CS gravity 
with that of a Schwarzschild black hole. Our results show a rapid increase in 
the deflection angle at small impact parameters upto a certain limit and then 
a gradual decrease is seen. As the deflection angle decreases, it is found to 
mimic the behaviour of the deflection angle for the Schawarschild black hole. 
At high impact parameters, the deflection angle for both the cases are found to 
overlap. The increase in the deflection angle at low impact parameter can be 
due to the presence of axonic hair, which is a result of the axion coupling 
with the curvature term. 

We obtain the Einstein ring radius and study its behaviour as a function of 
the impact parameter for the varying spin parameter and varying source 
distance. 
We also depict the change of the angular radius of the Einstein ring with 
change in the spin parameter. We see a gradual increase in the angular radius 
upto a certain limit and then it becomes constant for higher values of the 
impact parameter. Also, as the spin grows, the angular radius decreases. We 
can see that the Einstein ring radius is affected by the axionic hair of the 
slowly rotating black hole. The axionic hair also has some effect on the time 
delay of light. The time delay is seen to decrease with an increase in the 
spin of the black hole. As the source is assumed to be further away from the 
black hole, the time delay is seen to increase. This works comes with several 
futuristic outlooks. The magnification of the Einstein ring images formed 
around the black hole in CS gravity can be studied. Also, this work can be 
extended to obtain the deflection angle of light in wormhole backgrounds using 
different modified theories of gravity. 

In the subsequent stage of our investigation, we directed our attention 
towards the characteristics of the black hole shadow. We have derived the 
photon sphere radius and shadow expressions by utilising a perturbative scheme 
presuming the spin parameter $a$ to be sufficiently small enough. 
We have observed that the axionic coupling parameter 
$\eta$ is capable of producing an opposing influence to the black hole's spin, 
potentially obscuring the actual spin data from observational outcomes. The 
shadow of this black hole can be constrained by using Einstein Hubble 
Telescope data \cite{akiyama, EHT02} to have an observational constraint on the model parameters. We 
keep this as a future prospect of the study.

\section*{Acknowledgments}
UDG is thankful to the Inter-University Centre for Astronomy and Astrophysics
(IUCAA), Pune for the Visiting Associateship of the institute.


\begin{thebibliography}{99}

\bibitem{schneider}P. Schneider et al., {\em Gravitational lenses} (Berlin: Springer) (\href{https://doi.org/10.1007/978-3-662-03758-4}{https://doi.org/10.1007/978-3-662-03758-4}) (1992).

\bibitem{trimble}V. Trimble, The first lenses {\em Gravitational Lensing: Recent Progress and Future Goals} (ASP Conference Series vol 237) ed T G Brainerd and C S Kochanek (San Francisco, CA: Astronomical Society of the Pacific) pp 1-13 (2001).

\bibitem{renn}J. Renn et al., {\em The Origin of Gravitational Lensing: A Postscript to Einstein’s 1936 Science Paper}, \href{https://doi.org/10.1126/science.275.5297.184}{Science {\bf 275}, 184-6 (1997).}

\bibitem{valls}D. Valls-Gabaud, {\em The Conceptual Origins of Gravitational Lensing}, \href{https://doi.org/10.1063/1.2399715}{AIP Conf. Proc. {\bf 861}, 1163-71 (2006).}

\bibitem{eddington}F. W. Dyson et al., {\em IX. A Determination of the Deflection of Light by the Sun’s Gravitational Field, from Observations Made at the Total Eclipse of May 29, 1919}, \href{https://ui.adsabs.harvard.edu/link_gateway/1920RSPTA.220..291D/doi:10.1098/rsta.1920.0009}{Phil. Trans. R. Soc. A {\bf 220}, 291 (1920).} 

\bibitem{einstein}A. Einstein, {\em The Foundation of the General Theory of Relativity}, \href{https://doi.org/10.1002/andp.19163540702}{Ann. Phys. (N.Y.) {\bf 49}, 769 (1916), {\bf 14}, 517 (2005)}.

\bibitem{virbhadra}K. S. Virbhadra and G. F. R. Ellis, {\em Schwarzschild black hole lensing}, \href{https://doi.org/10.1103/PhysRevD.62.084003}{Phys. Rev. D {\bf 62}, 084003 (2000).}

\bibitem{zhao}F. Zhao et al., {\em Gravitational lensing effects of a Reissner–Nordstrom–de Sitter black hole}, \href{https://doi.org/10.1103/PhysRevD.93.123017}{Phys. Rev. D {\bf 93}, 123017 (2016).}

\bibitem{rindler}W. Rindler and M. Ishak, {\em Contribution of the cosmological constant to the relativistic bending of light revisited}, \href{https://doi.org/10.1103/PhysRevD.76.043006}{Phys. Rev. D {\bf 76}, 043006 (2007).}

\bibitem{ellis}K. S. Virbhadra and G. F. R. Ellis, {\em Gravitational lensing by naked singularities}, \href{https://doi.org/10.1103/PhysRevD.65.103004}{Phys. Rev. D {\bf 65}, 103004 (2002).}

\bibitem{shaikh}R. Shaikh et al., {\em Analytical approach to strong gravitational lensing from ultracompact objects}, \href{https://doi.org/10.1103/PhysRevD.99.104040}{Phys. Rev. D {\bf 99}, 104040 (2019).}

\bibitem{carmo}M. P. Do Carmo, {\textit {Differential geometry of curves andsurfaces}} (Dover Publications, Mineola, New York) (2016).

\bibitem{bonnet}W. Klingenberg, {\textit {A Course in Differential Geometry}} (Springer-Verlag, New York) (1978).

\bibitem{gibbons}G. W. Gibbons and M. C. Werner, {\em Applications of the Gauss-Bonnet theorem to gravitational lensing}, \href{https://doi.org/10.1088/0264-9381/25/23/235009}{Class. Quantum Grav. {\bf 25}, 235009 (2008).}

\bibitem{werner}M. C. Werner, {\em Gravitational Lensing in the Kerr-Randers Optical Geometry}, \href{https://doi.org/10.1007/s10714-012-1458-9}{Gen. Relativ. Gravit. {\bf 44}, 3047-3057 (2012)}.

\bibitem{saavedra}K. Jusufi et al., {\em Deflection of Light by Rotating Regular Black Holes Using the Gauss-Bonnet Theorem}, \href{https://doi.org/10.1103/PhysRevD.97.124024}{Phys. Rev. D {\bf 97}, 124024 (2018)}.

\bibitem{ishihara}A. Ishihara et al., {\em Gravitational Bending Angle of Light for Finite Distance and the Gauss-Bonnet Theorem}, \href{https://doi.org/10.1103/PhysRevD.94.084015}{Phys. Rev. D {\bf 94}, 084015 (2016)}.

\bibitem{ono}T. Ono, A. Ishihara and H. Asada, {\em Gravitomagnetic Bending Angle of Light with Finite-Distance Corrections in Stationary Axisymmetric Spacetimes}, \href{https://doi.org/10.1103/PhysRevD.96.104037}{Phys. Rev. D {\bf 96}, 104037 (2017)}.

\bibitem{kumar}R. Kumar, S. G. Ghosh and A. Wang, {\em Shadow Cast and Deflection of Light by Charged Rotating Regular Black Hole}, \href{https://doi.org/10.1103/PhysRevD.100.124024}{Phys. Rev. D {\bf 100}, 124024 (2019)}.

\bibitem{ghosh}R. Kumar, S. G. Ghosh and A. Wang, {\em Gravitational Deflection of Light and Shadow Cast by Rotating Kalb-Ramond Black Holes}, \href{https://doi.org/10.1103/PhysRevD.101.104001}{Phys. Rev. D {\bf 101}, 104001 (2020)}.

\bibitem{banerjee}K. Jusufi et al., {\em Light deflection by a rotating global monopole spacetime}, \href{https://doi.org/10.1103/PhysRevD.95.104012}{Phys. Rev. D {\bf 95}, 104012 (2017).} 

\bibitem{jusufi}K. Jusufi, {\em Gravitational lensing by Reissner-Nordström black holes with topological defects}, \href{https://doi.org/10.1007/s10509-015-2609-8}{Astrophys. Space Sci. {\bf 361}, 24 (2016).} 

\bibitem{kimet}K. Jusufi, {\em Light Deflection with Torsion Effects Caused by a Spinning Cosmic String}, \href{https://doi.org/10.1140/epjc/s10052-016-4185-7}{Eur. Phys. J. C {\bf 76}, 332 (2016)}.

\bibitem{sakalli}K. Jusufi, I. Sakalli and A. {\"O}vg{\"u}n, {\em Effect of Lorentz Symmetry Breaking on the Deflection of Light in a Cosmic String Spacetime}, \href{https://doi.org/10.1103/PhysRevD.96.024040}{Phys. Rev. D {\bf 96}, 024040 (2017)}.

\bibitem{Kjusufi}K. Jusufi, {\em Quantum Effects on the Deflection of Light and the Gauss-Bonnet Theorem}, \href{https://doi.org/10.1142/S0219887817501377}{Int. J. Geom. Methods Mod. Phys. {\bf 14}, 1750137 (2017)}.

\bibitem{izzet}I. Sakalli and A. {\"O}vg{\"u}n, {\em Hawking Radiation and Deflection of Light from Rindler Modified Schwarzschild Black Hole}, \href{https://doi.org/10.1209/0295-5075/118/60006}{EPL {\bf 118}, 60006 (2017)}.

\bibitem{ovgun1}A. {\"O}vg{\"u}n, G. Gyulchev and K. Jusufi, {\em Weak Gravitational Lensing by Phantom Black Holes and Phantom Wormholes Using the Gauss-Bonnet Theorem}, \href{https://doi.org/10.1016/j.aop.2019.04.007}{Ann. of Phys. {\bf 406}, 152-172 (2019)}.

\bibitem{ovgun2}A. {\"O}vg{\"u}n, {\em Weak Field Deflection Angle by Regular Black Holes with Cosmic Strings Using the Gauss-Bonnet Theorem}, \href{https://doi.org/10.1103/PhysRevD.99.104075}{Phys. Rev. D {\bf 99}, 104075 (2019)}.

\bibitem{ovgun3}Z. Li and A. {\"O}vg{\"u}n, {\em Finite-Distance Gravitational Deflection of Massive Particles by a Kerr-like Black Hole in the Bumblebee Gravity Model}, \href{https://doi.org/10.1103/PhysRevD.101.024040}{Phys. Rev. D {\bf 101}, 024040 (2020)}.

\bibitem{takizawa}K. Takizawa, T. Ono and H. Asada, {\em Gravitational Deflection Angle of Light: Definition by an Observer and Its Application to an Asymptotically Nonflat Spacetime}, \href{https://doi.org/10.1103/PhysRevD.101.104032}{Phys. Rev. D {\bf 101}, 104032 (2020)}.

\bibitem{crisnejo}G. Crisnejo, E. Gallo and K. Jusufi, {\em Higher Order Corrections to Deflection Angle of Massive Particles and Light Rays in Plasma Media for Stationary Spacetimes Using the Gauss-Bonnet Theorem}, \href{https://doi.org/10.1103/PhysRevD.100.104045}{Phys. Rev. D {\bf 100}, 104045 (2019)}.

\bibitem{pantig1}R. C. Pantig and A. {\"O}vg{\"u}n, {\em Dark Matter Effect on the Weak Deflection Angle by Black Holes at the Center of Milky Way and $M87$ Galaxies} \href{https://doi.org/10.1140/epjc/s10052-022-10319-8}{Eur. Phys. J. C {\bf 82}, 391 (2022)}.

\bibitem{pantig2}R. C. Pantig et al., {\em Shadow and Weak Deflection Angle of Extended Uncertainty Principle Black Hole Surrounded with Dark Matter}, \href{https://doi.org/10.1016/j.aop.2021.168722}{Ann. of Phys. {\bf 436}, 168722 (2022)}.

\bibitem{javed}W. Javed et al., {\em Weak Deflection Angle by Kalb-Ramond Traversable Wormhole in Plasma and Dark Matter Mediums}, \href{10.20944/preprints202209.0140.v1}{Preprints (2022)}.

\bibitem{ovgun4}A. {\"O}vg{\"u}n, {\em Weak Deflection Angle of Black-Bounce Traversable Wormholes UsingGauss–Bonnet Theorem in the Dark Matter Medium}, \href{https://doi.org/10.3906/fiz-2008-11}{Turk. J. Phys. {\bf 44}, 465-471 (2020)}.

\bibitem{bertone}G. Bertone and D. Hooper, {\em History of Dark Matter}, \href{https://doi.org/10.1103/RevModPhys.90.045002}{Rev. Mod. Phys. {\bf 90}, 045002 (2018).} 

\bibitem{swart}J. G. de Swart, G. Bertone and J. van Dongen, {\em How Dark Matter Came to Matter},\href{https://doi.org/10.1038/s41550-017-0059}{Nature Astron. {\bf 1}, 0059 (2017).} 

\bibitem{tim}G. Bertone and T. M. P. Tait, {\em A New Era in the Search for Dark Matter}, \href{https://doi.org/10.1038/s41586-018-0542-z}{Nature {\bf 562}, 51-56 (2018).} 

\bibitem{frenk}C. S. Frenk and S. D. M. White, {\em  Dark Matter and Cosmic Structure}, \href{https://dx.doi.org/10.1002/andp.201200212}{Ann. Phys. (Berlin), 507534 (2012).} 

\bibitem{strigari}L. E. Strigari, {\em Galactic Searches for Dark Matter}, \href{http://dx.doi.org/10.1016/j.physrep.2013.05.004}{Phys. Rep. {\bf 531}, 1-88 (2013).} 

\bibitem{will2014}C. M. Will, {\em The Confrontation between General Relativity and Experiment}, \href{https://doi.org/10.12942/lrr-2014-4}{Living Rev. Relativ. {\bf 17}, 4 (2014)}.

\bibitem{rubin}V. C. Rubin, N. Thonnard and W. K. Ford Jr., {\em Rotational properties of $21$ SC galaxies with a large range of luminosities and radii, from NGC $4605$ (R $= 4$ kpc) to UGC $2885$ (R $= 122$ kpc)}, \href{https://ui.adsabs.harvard.edu/link_gateway/1980ApJ...238..471R/doi:10.1086/158003}{Astrophys. J. {\bf 238}, 471-487 (1980)}.

\bibitem{young1}Bing-Lin Young, {\em A survey of dark matter and related topics in cosmology}, \href{https://doi.org/10.1007/s11467-016-0583-4}{Front. Phys. {\bf 12}, 121201 (2017)}.

\bibitem{harko}T. Harko, {\em Galactic rotation curves in modified gravity with nonminimal coupling between matter and geometry}, \href{http://dx.doi.org/10.1103/PhysRevD.81.084050}{Phys. Rev. D {\bf 81}, 084050 (2010)}.

\bibitem{gergely}L. \'{A}. Gergely et al., {\em Galactic rotation curves in brane world models}, \href{https://doi.org/10.1111/j.1365-2966.2011.18941.x}{Mon. Not. R. Astron. Soc. {\bf 415}, 3275-3290 (2011)}.

\bibitem{parbin}N. Parbin and U. D. Goswami, {\em Galactic rotation dynamics in a new $f(\mathcal{R})$ gravity model}, \href{https://doi.org/10.48550/arXiv.2208.06564}{[arXiv:2208.06564]}.

\bibitem{reiss}A. G. Reiss et al., {\em Observational Evidence from Supernovae for an Accelerating Universe and a Cosmological Constant}, \href{https://doi.org/10.1086/300499}{Astron. J. {\bf 116}, 1009 (1998).} 

\bibitem{perlmutter}S. Perlmutter et al., {\em Measurements of $\Omega$ and $\Lambda$ from 42 High-Redshift Supernovae}, \href{https://doi.org/10.1086/307221}{Astrophys. J. {\bf 517}, 565 (1999).}

\bibitem{Clifton}T. Clifton, {\em Alternative Theories of Gravity}, \href{https://arxiv.org/abs/gr-qc/0610071}{[arXiv:gr-qc/0610071] (2006).}

\bibitem{wagoner}R. V. Wagoner, {\em Scalar-Tensor Theory and Gravitational Waves}, \href{https://ui.adsabs.harvard.edu/link_gateway/1970PhRvD...1.3209W/doi:10.1103/PhysRevD.1.3209}{Phys. Rev. D {\bf 1}, 3209 (1970).}

\bibitem{ronald}R. W. Hellings and K. Nordtvedt, Jr., {\em Vector-Metric Theory of Gravity}, \href{https://ui.adsabs.harvard.edu/link_gateway/1973PhRvD...7.3593H/doi:10.1103/PhysRevD.7.3593}{Phys. Rev. D {\bf 7}, 3593 (1973).}

\bibitem{capozziello}S. Capozziello and M. De Laurentis, {\em Extended Theories of Gravity}, \href{http://dx.doi.org/10.1016/j.physrep.2011.09.003}{Phys. Rep. {\bf 509}, 167-321 (2011).} 

\bibitem{clifton}T. Clifton et al., {\em  Modified Gravity and Cosmology}, \href{http://dx.doi.org/10.1016/j.physrep.2012.01.001}{Phys. Rep. {\bf 513}, 1-189 (2012).} 

\bibitem{oikonomou}S. Nojiri, S.D. Odintsov and V.K. Oikonomou, {\em Modified Gravity Theories on a Nutshell: Inflation, Bounce and Late-Time Evolution}, \href{http://dx.doi.org/10.1016/j.physrep.2017.06.001}{Phys. Rept. {\bf 692}, 1-104 (2017).} 

\bibitem{odintsov}S. Nojiri and S.D. Odintsov, {\em Unified Cosmic History in Modified Gravity: From F(R) Theory to Lorentz Non-Invariant Models}, \href{http://dx.doi.org/10.1016/j.physrep.2011.04.001}{Phys. Rept. {\bf 505}, 59-144 (2011).} 

\bibitem{sergei}S. Nojiri and S.D. Odintsov, {\em Dark Energy, Inflation and Dark Matter from Modified F(R) Gravity}, TSPU Bulletin N {\bf 8(110)}, 7 (2011) \href{https://arxiv.org/abs/0807.0685}{[arXiv:0807.0685]}.

\bibitem{gogoi1}D. J. Gogoi and U. D. Goswami, {\em A New f(R) Gravity Model and Properties of Gravitational Waves in It}, \href{https://doi.org/10.1140/epjc/s10052-020-08684-3}{Eur. Phys. J. C {\bf 80}, 1101 (2020).}

\bibitem{gogoi_cosmo}D. J. Gogoi and U. D. Goswami, {\em Cosmology with a new f(R) gravity model in Palatini formalism}, \href{https://doi.org/10.1142/S0218271822500481}{Int. J. Mod. Phys. D 31, 2250048 (2022).}

\bibitem{nashiba}N. Parbin and U. D. Goswami, {\em Scalarons Mimicking Dark Matter in the Hu-Sawicki Model of f(R) Gravity}, \href{https://dx.doi.org/10.1142/S0217732321502655}{Mod. Phys. Lett. A {\bf 36}, 2150265 (2021)}.

\bibitem{dorlis}N. Chatzifotis, et al., {\em Scalarization of Chern-Simons-Kerr Black Hole Solutions and Wormholes}, \href{https://doi.org/10.1103/PhysRevD.105.084051}{Phys. Rev. D {\bf 105}, 084051 (2022)}.

\bibitem{jackiw}R. Jackiw and S.-Y. Pi, {\em  Chern-Simons Modification of General Relativity}, \href{https://doi.org/10.1103/PhysRevD.68.104012}{Phys. Rev. D {\bf 68}, 104012 (2003)}.

\bibitem{konno}K. Konno et al., {\em Flat Rotation Curves in Chern-Simons Modified Gravity}, \href{http://dx.doi.org/10.1103/PhysRevD.78.024037}{Phys. Rev. D {\bf 78}, 024037 (2008)}.

\bibitem{corral}C. Corral, C. Erices, D. Flores-Alfonso, and K. Lara, {\em Phase Transitions of Black Strings in Dynamical Chern-Simons Modified Gravity}, \href{https://doi.org/10.1103/PhysRevD.105.024050}{Phys. Rev. D {\bf 105}, 024050 (2022)}.

\bibitem{doneva}D. D. Doneva and S. S. Yazadjiev, {\em Spontaneously Scalarized Black Holes in Dynamical Chern-Simons Gravity: Dynamics and Equilibrium Solutions}, \href{https://doi.org/10.1103/PhysRevD.103.083007}{Phys. Rev. D {\bf 103}, 083007 (2021)}.

\bibitem{shaoqi}S. Hou, T. Zhu and Z.-H. Zhu, {\em Asymptotic Analysis of Chern-Simons Modified Gravity and Its Memory Effects}, \href{https://doi.org/10.1103/PhysRevD.105.024025}{Phys. Rev. D {\bf 105}, 024025 (2022)}.

\bibitem{tiberiu}T. Harko, Z. Kov{\'a}cs and F. S. N. Lobo, {\em Thin Accretion Disk Signatures in Dynamical Chern-Simons-Modified Gravity}, \href{http://dx.doi.org/10.1088/0264-9381/27/10/105010}{Class. Quantum Grav. {\bf 27}, 105010 (2010)}.

\bibitem{popov}S. Nojiri et al., {\em Propagation of Gravitational Waves in Chern-Simons Axion Einstein Gravity}, \href{https://doi.org/10.1103/PhysRevD.100.084009}{Phys. Rev. D {\bf 100}, 084009 (2019)}.

\bibitem{cisterna}A. Cisterna, C. Corral and S. del Pino, {\em Static and Rotating Black Strings in Dynamical Chern–Simons Modified Gravity}, \href{https://doi.org/10.1140/epjc/s10052-019-6910-5}{Eur. Phys. J. C {\bf 79}, 400 (2019)}.

\bibitem{matsuyama}K. Konno, T. Matsuyama, and S. Tanda, {\em Does a Black Hole Rotate in Chern-Simons Modified Gravity?}, \href{http://dx.doi.org/10.1103/PhysRevD.76.024009}{Phys. Rev. D {\bf 76}, 024009 (2007)}.

\bibitem{myung}T. Moon and Y. S. Myung, {\em Stability of the Schwarzschild Black Hole in $f(R)$ Gravity with the Dynamical Chern-Simons Term}, \href{http://dx.doi.org/10.1103/PhysRevD.84.104029}{Phys. Rev. D {\bf 84}, 104029 (2011)}.

\bibitem{molina}C. Molina et al., {\em Gravitational Signature of Schwarzschild Black Holes in Dynamical Chern-Simons Gravity}, \href{http://dx.doi.org/10.1103/PhysRevD.81.124021}{Phys. Rev. D {\bf 81}, 124021 (2010)}.

\bibitem{cardoso}V. Cardoso and L. Gualtieri, {\em Perturbations of Schwarzschild Black Holes in Dynamical Chern-Simons Modified Gravity}, \href{http://dx.doi.org/10.1103/PhysRevD.80.064008}{Phys. Rev. D {\bf 80}, 064008 (2009)}.

\bibitem{witten}P. Svrcek and E. Witten, {\em Axions in String Theory}, \href{https://doi.org/10.1088/1126-6708/2006/06/051}{J. High Energy Phys. {\bf 06}, 051 (2006)}.

\bibitem{wonwoo}B.-H. Lee, W. Lee, and Y. S. Myung, {\em Shadow Cast by a Rotating Black Hole with Anisotropic Matter}, \href{https://doi.org/10.1103/PhysRevD.103.064026}{Phys. Rev. D {\bf 103}, 064026 (2021)}.

\bibitem{gussmann}A. Gu{\ss}mann, Polarimetric Signatures of the Photon Ring of a Black Hole That Is Pierced by a Cosmic Axion String, \href{https://doi.org/10.1007/JHEP08(2021)160}{J. High Energ. Phys. {\bf 2021}, 160 (2021)}.



\bibitem{jamil}T. Zhu et al., {\em Shadows and Deflection Angle of Charged and Slowly Rotating Black Holes in Einstein-{\AE}ther Theory}, \href{https://doi.org/10.1103/PhysRevD.100.044055}{Phys. Rev. D {\bf 100}, 044055 (2019)}.



\bibitem{ovgun5}A. {\"O}vg{\"u}n, I. Sakalli and J. Saavedra, {\em Shadow Cast and Deflection Angle of Kerr-Newman-Kasuya Spacetime}, \href{https://doi.org/10.1088/1475-7516/2018/10/041}{J. Cosmol. Astropart. Phys. {\bf 10}, 041 (2018)}.

\bibitem{wei}S.-W. Wei and Y.-X. Liu, {\em Observing the Shadow of Einstein-Maxwell-Dilaton-Axion Black Hole}, \href{https://doi.org/10.1088/1475-7516/2013/11/063}{J. Cosmol. Astropart. Phys. {\bf 11}, 063 (2013)}.

\bibitem{cunha}P. V. P. Cunha et al., {\em Shadows of Einstein–Dilaton–Gauss–Bonnet Black Holes}, \href{https://doi.org/10.1016/j.physletb.2017.03.020}{Physics Letters B {\bf 768}, 373 (2017)}.

\bibitem{wang}
H.-M. Wang, Y.-M. Xu and S.-W. Wei, {\em Shadows of Kerr-like Black Holes in a Modified Gravity Theory}, \href{https://doi.org/10.1088/1475-7516/2019/03/046}{J. Cosmol. Astropart. Phys. {\bf 03}, 046 (2019)}.

\bibitem{dastan}S. Dastan, R. Saffari and S. Soroushfar, {\em Shadow of a Charged Rotating Black Hole in $f(R)$ Gravity}, \href{https://doi.org/10.1140/epjp/s13360-022-03218-0}{Eur. Phys. J. Plus {\bf 137}, 1002 (2022)}.

\bibitem{1s}
K. Jusufi, \textit{Quasinormal Modes of Black Holes Surrounded by Dark Matter and Their Connection with the Shadow Radius}, \href{https://doi.org/10.1103/PhysRevD.101.084055}{Phys. Rev. D \textbf{101}, 084055 (2020).}

\bibitem{shnew02} R. C. Pantig, L. Mastrototaro, G. Lambiase, and A. \"Ovg\"un, {\it Shadow, Lensing, Quasinormal Modes, Greybody Bounds and Neutrino Propagation by Dyonic ModMax Black Holes}, \href{https://doi.org/10.1140/epjc/s10052-022-11125-y}{Eur. Phys. J. C 82, 1155 (2022).}


\bibitem{shnew01} {\.I}. {\c C}imdiker, D. Demir, and A. \"Ovg\"un, {\it Black Hole Shadow in Symmergent Gravity}, \href{
https://doi.org/10.1016/j.dark.2021.100900}{Physics of the Dark Universe {\bf 34}, 100900 (2021).}


\bibitem{2s}
Z. Li and C. Bambi, \textit{Measuring the Kerr spin parameter of regular black holes from their shadow}, \href{https://doi.org/10.1103/PhysRevD.80.024042}{JCAP 1401, 041 (2014).}

\bibitem{4s}
C. Bambi and K. Freese, \textit{Apparent shape of super-spinning black holes}, \href{https://doi.org/10.1103/PhysRevD.79.043002}{Phys. Rev. D 7\textbf{9}, 043002 (2009).}

\bibitem{5s}
S. Haroon, K. Jusufi and M. Jamil, \textit{Shadow Images of a Rotating Dyonic Black Hole with a Global Monopole Surrounded by Perfect Fluid}, \href{https://www.mdpi.com/2218-1997/6/2/23}{Universe \textbf{6} (2),23 (2020).}

\bibitem{6s}
M. Okyay and A. \"{O}vg\"{u}n, \textit{Nonlinear electrodynamics effects on the black hole shadow, deflection angle, quasinormal modes and greybody factors}, \href{https://iopscience.iop.org/article/10.1088/1475-7516/2022/01/009}{JCAP \textbf{01}, 009 (2022).}

\bibitem{7s}
A. Belhaj and Y. Sekhmani, \textit{Shadows of rotating quintessential black holes in Einstein–Gauss–Bonnet gravity with a cloud of strings}, \href{https://link.springer.com/article/10.1007/s10714-022-02902-x}{Gen. Relativ Gravit. \textbf{54} (2021).}

\bibitem{8s}
A. Allahyari, M. Khodadi, S. Vagnozzi and D. F. Mota, \textit{Magnetically charged black holes from non-linear electrodynamics and the Event Horizon Telescope}, \href{https://iopscience.iop.org/article/10.1088/1475-7516/2020/02/003}{JCAP \textbf{02}, 003 (2020).}

\bibitem{9s}
C. Bambi, K. Freese, S. Vagnozzi and L. Visinelli, \textit{Testing the rotational nature of supermassive object $M87^*$ from the circularity and size of its first image}, \href{https://journals.aps.org/prd/abstract/10.1103/PhysRevD.100.044057}{Phys. Rev. D \textbf{100}, 044057 (2019).}

\bibitem{10s}
S. Vagnozzi, C. Bambi and L. Visinelli, \textit{Concerns regarding the use of black hole shadows as standard rulers}, \href{https://iopscience.iop.org/article/10.1088/1361-6382/ab7965}{Class. Quantum Grav. \textbf{37}, 087001 (2020).}

\bibitem{11s}
M. Khodadi, A. Allahyari, S. Vagnozzi and D. F. Mota, \textit{Black holes with scalar hair in light of the Event Horizon Telescope}, \href{https://iopscience.iop.org/article/10.1088/1475-7516/2020/09/026}{JCAP \textbf{09}, 026 (2020).}

\bibitem{12s}
R. Roy, S. Vagnozzi and L. Visinelli, \textit{Superradiance evolution of black hole shadows revisited}, \href{https://journals.aps.org/prd/abstract/10.1103/PhysRevD.105.083002}{Phys. Rev. D \textbf{105}, 083002 (2022).}


\bibitem{13s}
B. E. Panah, Kh. Jafarzade and A. Rincon, \textit{Three-dimensional AdS black holes in massive-power-Maxwell theory}, \href{https://doi.org/10.48550/arXiv.2201.13211}{[arXiv:2201.13211v1] (2022).}


\bibitem{14s}
R. Ghosh, M. Rahman and A. K. Mishra, \textit{Regularized Stable Kerr Black Hole: Cosmic Censorships, Shadow and Quasi-Normal Modes}, \href{https://arxiv.org/abs/2209.12291}{arXiv:2209.12291 [gr-qc] (2023).}

\bibitem{15s}
R. A. Konoplya, \textit{Shadow of a black hole surrounded by dark matter}, \href{https://www.sciencedirect.com/science/article/pii/S0370269319303648?via%3Dihub}{Phys. Lett B \textbf{795}, 1-6 (2019).}

\bibitem{16s}
R. A. Konoplya and A. Zhidenko, \textit{Shadows of parametrized axially symmetric black holes allowing for separation of variables}, \href{https://journals.aps.org/prd/abstract/10.1103/PhysRevD.103.104033}{Phys. Rev. D \textbf{103}, 104033 (2021).}

\bibitem{17s}
K. Jusufi and Saurabh, \textit{ Black hole shadows in Verlinde’s emergent gravity}, \href{https://academic.oup.com/mnras/article-abstract/503/1/1310/6144591?redirectedFrom=fulltext&login=true}{MNRAS \textbf{503}, 1310 (2021).}

\bibitem{18s}
T. Zhu, Q. Wu, M. Jamil and K. Jusufi, \textit{Shadows and deflection angle of charged and slowly rotating black holes in Einstein-Æther theory}, \href{https://journals.aps.org/prd/abstract/10.1103/PhysRevD.100.044055}{Phys. Rev. D \textbf{100}, 044055 (2019).}

\bibitem{19s}
K. Jusufi et. al., \textit{Black hole surrounded by a dark matter halo in the M87 galactic center and its identification with shadow images}, \href{https://journals.aps.org/prd/abstract/10.1103/PhysRevD.100.044012}{Phys. Rev. D \textbf{100}, 044012 (2019).}



\bibitem{102-1}B. Mclnnes and Y. C. Ong, \textit{Event horizon wrinklification}, \href{https://iopscience.iop.org/article/10.1088/1361-6382/abce45}{Class. quantum Grav. \textbf{38}, 034002 (2021).}

\bibitem{102-2}M. Khodadi, E. N. Saridakis, \textit{Einstein-{\AE}ther gravity in the light of event horizon telescope observations of M87*}, \href{https://www.sciencedirect.com/science/article/abs/pii/S2212686421000662?via%3Dihub}{Physics of the Dark Universe \textbf{32}, 100835 (2021).}

\bibitem{102-3}K. Glampedakis and G. Pappas, \textit{Can supermassive black hole shadows test the Kerr metric?}, \href{https://journals.aps.org/prd/abstract/10.1103/PhysRevD.104.L081503}{Phys. Rev. D \textbf{104}, L081503 (2021).}

\bibitem{102-4}S. Devi, A. Nagarajan S., S. Chakrabarty and B. R. Majhi, \textit{Shadow of quantum extended Kruskal black hole and its super-radiance property}, \href{https://www.sciencedirect.com/science/article/abs/pii/S2212686423000079?via%3Dihub}{Physics of the Dark Universe \textbf{39}, 101173 (2023).}

\bibitem{102-5}M. Khodadi, G. Lambiase and D. Mota, \textit{No-hair theorem in the wake of Event Horizon Telescope}, \href{https://iopscience.iop.org/article/10.1088/1475-7516/2021/09/028}{JCAP \textbf{09} 028 (2021).}

\bibitem{102-6}M. Khodadi and G. Lambiase, \textit{Probing Lorentz symmetry violation using the first image of Sagittarius A*: Constraints on standard-model extension coefficients}, \href{https://journals.aps.org/prd/abstract/10.1103/PhysRevD.106.104050}{Phys. Rev. D \textbf{106}, 104050 (2022).}

\bibitem{102-7}I. Banerjee, S. Chakrabarty and S. SenGupta, \textit{Hunting extra dimensions in the shadow of Sgr A*}, \href{https://journals.aps.org/prd/abstract/10.1103/PhysRevD.106.084051}{Phys. Rev. D \textbf{106}, 084051 (2022).}

\bibitem{102-8}M. Afrin, S. Vagnozzi and S. G. Ghosh, \textit{Tests of Loop Quantum Gravity from the Event Horizon Telescope Results of Sgr A*}, \href{https://iopscience.iop.org/article/10.3847/1538-4357/acb334}{APJ \textbf{944}, 2 (2023).}

\bibitem{102-9}M. Khodadi, \textit{Shadow of black hole surrounded by magnetized plasma: Axion-plasmon cloud}, \href{https://www.sciencedirect.com/science/article/pii/S0550321322003650?via%3Dihub}{Nuclear Phys. B \textbf{985}, 116014 (2022).}

\bibitem{102-10}S. Chen, J. Jing, W. L. Qian and B. Wang, \textit{Black hole images: A Review}, \href{https://link.springer.com/article/10.1007/s11433-022-2059-5#citeas}{Sci. China Phys. Mech. Astron. \textbf{66}, 260401 (2023).}

\bibitem{102-11}S. Vagnozzi and L. Visinelli, \textit{Hunting for extra dimensions in the shadow of M87*}, \href{https://journals.aps.org/prd/abstract/10.1103/PhysRevD.100.024020}{Phys. Rev. D \textbf{100}, 024020 (2019).}

\bibitem{102-12}Y. Chen, R. Roy, S. Vagnozzi and L. Visinelli, \textit{Superradiant evolution of the shadow and photon ring of Sgr A*}, \href{https://journals.aps.org/prd/abstract/10.1103/PhysRevD.106.043021}{Phys. Rev. D \textbf{106}, 043021 (2022).}

\bibitem{102-13}S. Vagnozzi et. al., \textit{Horizon-scale tests of gravity theories and fundamental physics from the Event Horizon Telescope image of Sagittarius A*}, \href{https://doi.org/10.48550/arXiv.2205.07787}{arXiv:2205.07787 [gr-qc] (2022).}

\bibitem{102-14}Y. Hou, M. Guo and B. Chen, \textit{Revisiting the shadow of braneworld black holes}, \href{https://journals.aps.org/prd/abstract/10.1103/PhysRevD.104.024001}{Phys. Rev D \textbf{104}, 024001 (2021).}

\bibitem{102-15}P. Li, M. Guo and B. Chen, \textit{Shadow of a spinning black hole in an expanding universe}, \href{https://journals.aps.org/prd/abstract/10.1103/PhysRevD.101.084041}{phys. Rev. D \textbf{101}, 084041 (2020).}

\bibitem{3s}
G. Gyulchev, P. Nedkova, V. Tinchev and S. Yazadjiev, \textit{On the shadow of rotating traversable wormholes}, \href{https://link.springer.com/article/10.1140/epjc/s10052-018-6012-9}{Eur. Phys. J. C \textbf{78} (7), 544 (2018).}

\bibitem{akiyama}
Event Horizon Telescope Collaboration et al., {\em First M87 Event Horizon Telescope Results. IV. Imaging the Central Supermassive Black Hole}, ApJ {\bf 875}, L4 (2019).

\bibitem{carvalho}I. D. D. Carvalho et al., {\em The Gravitational Bending Angle by Static and Spherically Symmetric Black Holes in Bumblebee Gravity} [\href{https://doi.org/10.48550/arXiv.2103.03845}{arXiv:2103.03845}].

\bibitem{ronit23} R. Karmakar, D. J. Gogoi, and U. D. Goswami, {\it Thermodynamics and Shadows of GUP-Corrected Black Holes with Topological Defects in Bumblebee Gravity}, \href{https://doi.org/10.1016/j.dark.2023.101249}{Physics of the Dark Universe {\bf 41}, 101249 (2023).}

\bibitem{parbin23} N. Parbin, D. J. Gogoi, J. Bora, and U. D. Goswami, {\it Deflection Angle, Quasinormal Modes and Optical Properties of a de Sitter Black Hole in $f(\mathcal{T}, \mathcal{B})$ Gravity}, \href{https://arxiv.org/abs/2211.02414}{arXiv:2211.02414} (2023).

\bibitem{hou}X. Hou et al., {\em Black Hole Shadow of Sgr A$^*$ in Dark Matter Halo}, \href{https://doi.org/10.1088/1475-7516/2018/07/015}{J. Cosmol. Astropart. Phys. {\bf 07}, 015 (2018)}.

\bibitem{konoplya}R. A. Konoplya, {\em Shadow of a Black Hole Surrounded by Dark Matter}, \href{https://doi.org/10.1016/j.physletb.2019.05.043}{Physics Letters B {\bf 795}, 1 (2019)}.

\bibitem{haroon}S. Haroon et al., {\em Shadow and Deflection Angle of Rotating Black Holes in Perfect Fluid Dark Matter with a Cosmological Constant}, \href{https://doi.org/10.1103/PhysRevD.99.044015}{Phys. Rev. D {\bf 99}, 044015 (2019)}.

\bibitem{salucci}K. Jusufi et al., {\em Black Hole Surrounded by a Dark Matter Halo in the M87 Galactic Center and Its Identification with Shadow Images}, \href{https://doi.org/10.1103/PhysRevD.100.044012}{Phys. Rev. D {\bf 100}, 044012 (2019)}.

\bibitem{barkana}W. Hu, R. Barkana and A. Gruzinov, {\em Fuzzy Cold Dark Matter: The Wave Properties of Ultralight Particles}, \href{https://doi.org/10.1103/PhysRevLett.85.1158}{Phys. Rev. Lett. {\bf 85}, 1158 (2000)}.

\bibitem{cicoli}M. Cicoli et al., {\em Fuzzy Dark Matter Candidates from String Theory}, \href{https://doi.org/10.1007/JHEP05(2022)107}{J. High Energ. Phys. {\bf 2022}, 107 (2022)}.

\bibitem{carosi}J. E. Kim and G. Carosi, {\em Axions and the Strong C P Problem}, \href{https://doi.org/10.1103/RevModPhys.82.557}{Rev. Mod. Phys. {\bf 82}, 557 (2010)}; Erratum \href{https://journals.aps.org/rmp/abstract/10.1103/RevModPhys.91.049902}{Rev. Mod. Phys. {\bf 91}, 049902 (2019)}.

\bibitem{ringwald}A. Ringwald, {\em Exploring the Role of Axions and Other WISPs in the Dark Universe}, \href{https://doi.org/10.1016/j.dark.2012.10.008}{Physics of the Dark Universe {\bf 1}, 116 (2012)}.

\bibitem{kawasaki}M. Kawasaki and K. Nakayama, {\em Axions: Theory and Cosmological Role}, \href{https://doi.org/10.1146/annurev-nucl-102212-170536}{Annu. Rev. Nucl. Part. Sci. {\bf 63}, 69 (2013)}.

\bibitem{yunes}S. Alexander and N. Yunes, {\em Chern-Simons Modified General Relativity}, \href{https://doi.org/10.1016/j.physrep.2009.07.002}{Physics Reports {\bf 480}, 1 (2009)}.

\bibitem{duncan}M. J. Duncan, N. Kaloper and K. A. Olive, {\em Axion Hair and Dynamical Torsion from Anomalies}, \href{https://doi.org/10.1016/0550-3213(92)90052-D}{Nuclear Physics B {\bf 387}, 215 (1992)}.

\bibitem{basilakos}S. Basilakos, N. E. Mavromatos and J. Solà Peracaula, {\em Gravitational and Chiral Anomalies in the Running Vacuum Universe and Matter-Antimatter Asymmetry}, \href{https://doi.org/10.1103/PhysRevD.101.045001}{Phys. Rev. D {\bf 101}, 045001 (2020)}.

\bibitem{bozza}
V. Bozza, {\em Comparison of Approximate Gravitational Lens Equations and a Proposal for an Improved New One}, \href{http://dx.doi.org/10.1103/PhysRevD.78.103005}{Phys. Rev. D {\bf 78}, 103005 (2008)}.

\bibitem{papnoi}
F. Atamurotov, U. Papnoi, and K. Jusufi, {\em Shadow and Deflection Angle of Charged Rotating Black Hole Surrounded by Perfect Fluid Dark Matter}, \href{https://doi.org/10.1088/1361-6382/ac3e76}{Class. Quantum Grav. {\bf 39}, 025014 (2022)}.


\bibitem{weinberg}S. Weinberg, {\em Gravitation and Cosmology: Principles and Applications of the General Theory of Relativity} (Wiley, New York, 1972).

\bibitem{keeton}C. R. Keeton and A. O. Petters, {\em Formalism for Testing Theories of Gravity Using Lensing by Compact Objects: Static, Spherically Symmetric Case}, \href{http://dx.doi.org/10.1103/PhysRevD.72.104006}{Phys. Rev. D {\bf 72}, 104006 (2005)}.

\bibitem{mancini}V. Bozza and L. Mancini, {\em Time Delay in Black Hole Gravitational Lensing as a Distance Estimator}, \href{https://doi.org/10.1023/B:GERG.0000010486.58026.4f}{Gen. Rel. Grav. {\bf 36}, 435 (2004)}.

\bibitem{adair} C. Adair, P. Bueno, P. A. Cano, R. A. Hennigar, and R. B. Mann, Slowly Rotating Black Holes in Einsteinian Cubic Gravity, \href{https://doi.org/10.1103/PhysRevD.102.084001}{Phys. Rev. D {\bf 102}, 084001 (2020).}

\bibitem{Gammon}
M. Gammon and R. Mann, {\it Slowly Rotating Black Holes in 4D Gauss-Bonnet Gravity}, 	\href{https://arxiv.org/abs/2210.01909}{arXiv:2210.01909} (2022).


\bibitem{EHT02}K. Akiyama et al. (Event Horizon Telescope), {\it First Sagittarius A* Event Horizon Telescope Results. I. The Shadow of the
Supermassive Black Hole in the Center of the Milky Way}, \href{https://doi.org/10.3847/2041-8213/ac6674}{Astrophys. J. Lett. {\bf 930}, L12 (2022)}.

\end{thebibliography}
\end{document}